\documentclass[%
preprint, 
nofootinbib,
 amsmath,amssymb,
 aps, physrev,
]{revtex4-2}

\usepackage[dvipsnames]{xcolor}
\usepackage{comment}
\usepackage{cancel}
\usepackage{subcaption}
\usepackage{makecell}
\usepackage{ulem}
\usepackage{multirow}
\usepackage{rotating}
\usepackage{booktabs,makecell,siunitx,rotating}

\usepackage{graphicx}
\usepackage{dcolumn}
\usepackage{bm}

\newcommand{\tcg}[1]{\textcolor{ForestGreen}{#1}}
\newcommand{\tcb}[1]{\textcolor{blue}{#1}}
\newcommand{\tcr}[1]{\textcolor{red}{#1}}

\begin{document}

\preprint{APS/123-QED}

\title{\textbf{Demonstration of an integral method for estimating\\wall shear stress in complex high-speed flows} 
}%

\author{Mateus A. R. Braga}
 \email{Contact author: mateus.braga@colorado.edu}
\author{Robyn L. Macdonald}%
\affiliation{%
 University of Colorado Boulder\\Ann \& H.J. Smead Department of Aerospace Engineering Sciences\\3775 Discovery Dr., Boulder, CO 80303, USA
}%

\date{\today}

\begin{abstract}
Turbulent flows over blunt bodies with distributed roughness present a class of problems relevant to hypersonic atmospheric entry systems. However, accurate predictions of shear stress on such bodies remains elusive. This work presents a simple integral formulation to infer wall shear stress based on the Favre-averaged streamwise momentum equation, integrated once in the wall-normal direction. The proposed integral formulation eliminates streamwise dependence, relying only on data and gradients extracted in the wall normal direction. Eight demonstration cases were selected to show the contributions of the various terms of the integral equation, the associated error in the estimate, and outline practical considerations when estimating the wall shear stress for complex flow conditions. In all cases, the error in the predicted shear stress compared to a more traditional approach was no more than 5\%, with many cases having much lower error. Notably, the method was found to be viable even for surfaces in the transitionally rough and fully rough regimes by appropriate selection of a virtual origin, as well as flows over curved surfaces or with pressure gradients. Finally, the method produces acceptable estimates of shear stress even in the extreme condition when over 40\% of the near-wall boundary layer data is absent. In brief, the present integral method is general and applicable to flows with curvature, surface roughness, and pressure gradients.

\end{abstract}

\maketitle


\newcommand{\pp}[2]{\frac{\partial #1}{\partial #2}}

\section{\label{sec:Introduction}Introduction}
Accurately estimating the wall shear stress is paramount for the analysis and characterization of wall-bounded fluid flows. A multitude of techniques exist in the literature to compute the wall shear stress, yet no one method is adequate for every condition. One such challenging condition is hypersonic flow with distributed surface roughness. In general, atmospheric entry systems make use of blunt body geometries and ablative thermal protection systems which introduce distributed or patterned roughness to the surface. This work formalizes a simple integral formulation to compute the wall shear stress from the Favre-averaged streamwise momentum equation. The approach is general enough to be applicable regardless of streamwise gradients and surface roughness.

Experimental measurement techniques exist to measure the drag force per unit area exerted by a fluid on a surface. Techniques include floating-element force balance sensors, oil-film interferometry, Preston tubes, and Stanton Tubes to name a few. An overview of experimental shear stress measurement techniques can be found in \cite{Vinuesa17}. Correlations are also common; a recent example from Dixit et al. \cite{Dixit24} showed a generalized scaling framework and predictive model for wall friction in turbulent flow based on a variety of datasets, including compressible flow simulations. However, there were no rough-walled cases in the data fit and only two cases above Mach 1 (one supersonic and one hypersonic). Moreover, correlations are generally limited to conditions similar to those used to construct the fit and should be used with caution outside such conditions. Although there are benefits and limitations to the direct measurement techniques and generalized scalings, they will not be discussed further, and the focus of this work is on inferred shear stress estimates when wall-normal velocity, density, and pressure profiles are available, obtained experimentally or through computer simulation. 

To highlight the need for the wall shear stress estimation approach employed in this work, a brief description of some of the most common methods to infer the wall shear stress is provided. By definition, the shear stress $\tau_w$ is the manifestation of the loss of streamwise momentum in the fluid due to the presence of the surface. For Newtonian fluids over smooth walls, the shear stress at a wall is $\tau_w=\mu\pp{u}{y}$, where $\tau_w$ is the wall shear stress, $\mu$ is the fluid dynamic viscosity, $u$ is the streamwise velocity, and $y$ is the wall-normal coordinate. In cases where the velocity profile in the viscous sublayer is known, the shear stress may be computed directly from the aforementioned definition. In the presence of surface roughness, the form drag becomes significant \cite{Perry69}, and the estimate of the shear stress should also consider the net effect of the surface pressure distribution. The pressure contribution may be computed from the surface integral $\frac{1}{A}\int_Sp_w\hat{n}\cdot \hat{e}_xdS$, where $A$ is the projected planform area, $p_w$ is the surface pressure, $\hat{n}$ is the wall-normal unit vector, and $\hat{e}_x$ is the streamwise unit vector. Unfortunately, accurate measurements of the near-wall velocity can be difficult to obtain, either through the limitations of experimental measurement techniques or in the presence of surface roughness. Similarly, full surface pressure distributions with resolution adequate for integration are often limited to scale-resolving simulations like direct numerical simulation (DNS) or wall-resolved large eddy simulation (WRLES). For fully developed channel flow, a preferred method is to compute the pressure drop across the channel and relate it to the local tangential force (per area) exerted by the fluid~\cite{Pope00,Kuwata22}. This method is applicable to flows with or without roughness in incompressible or compressible regimes.

For wall-bounded turbulent flows, another method that infers the shear stress solely from the mean velocity profile is the Clauser fit technique \cite{Clauser54} which seeks a value of the friction velocity, $u_\tau$, such that the inner scaled mean velocity profile fits the overlap region's logarithmic law (log-law) $u^+ = (1/\kappa)\ln\left(y^+\right)+C$. The friction velocity is selected based on the best agreement between the log-law and the inner-scaled mean velocity for $y^+ > 50$ \cite{Vinuesa17}. Classically, the inner-scaled mean velocity is given by $u^+=\overline{u}/u_\tau$, where the dimensional mean velocity is $\overline{u}$, the inner-scaled wall-normal coordinate is $y^+=(yu_\tau)/\nu$, the dimensional wall-normal coordinate is $y$, and the kinematic viscosity is $\nu$. For compressible flow, Van Driest \cite{VanDriest1951} or Trettel and Larsson \cite{Trettel2016} transformations should be used. The Clauser fit is a common technique and only requires one wall-normal velocity profile at the location of interest. This method has been shown to work for a variety of surface conditions such as fully-developed, zero-pressure-gradient (ZPG) smooth-wall or rough-wall boundary layers \cite{Clauser54,Kong2024,Rowin2024}. However, in cases where the universal log-law does not hold, such as with strong pressure gradients, this method becomes inaccurate \cite{Dixit2009,Butt2018}   

Another well established method of determining the wall shear is to derive an expression for the wall shear stress (or skin friction coefficient) from a form of the momentum integral equation. This has been well studied for incompressible flows \cite{Johansson02, Fukagata02, Brzek07, Mehdi11, Mehdi14, Deck14, Yoon16, Renard16, Volino18, Ricco22, Elnahhas22, Li23}. Much of the effort has gone into various formulations that permit accurate estimates of the wall shear stress or skin friction coefficient without the need for streamwise gradients or near-wall data which may be hard to acquire experimentally or in the presence of roughness. Recent work has extended these various formulations to compressible flows. For example, Li et al. \cite{Li19} extend from Renard and Deck \cite{Renard16}, developing a compressible form decomposing the contributions to the shear stress into two components, namely the power of the friction transformed into heat via direct molecular viscous dissipation, and the power converted into turbulent kinetic-energy production. The results from Renard and Deck \cite{Renard16} or Li et al. \cite{Li19} focused on zero-pressure-gradient boundary layers or for channel flow. Passiatore et al. \cite{Passiatore22} subsequently uses Li et al. \cite{Li19} in the context of a hypersonic Mach 12.48 turbulent boundary layer. Xu et al. \cite{Xu23} derive an integral formula for the skin friction coefficient of compressible boundary layers by extending the formula of Elnahhas and Johnson \cite{Elnahhas22} that was originally for incompressible flow. One of the key findings by Xu et al. \cite{Xu23} was confirmation that the integral identities with twofold and threefold integrations can be sensitive to the selection of the upper bound of integration. They sought an identity that removes the dependence on the upper bound of integration and is, therefore, valid for compressible boundary layers with an unbounded domain in the wall-normal direction. Their decomposition is subsequently applied to numerical data for laminar and turbulent (flat plate) boundary layers, where the role of each term on the skin friction coefficient is quantified. 

The present work aims to build off these integral formulations to demonstrate the capability of integral methods for estimating wall shear stress in complex high-speed flows, such as with surface roughness, or surface curvature. Starting from the Favre-averaged streamwise momentum equation, we integrate in the wall-normal direction only once to obtain a stress balance from which the wall-shear stress can be inferred. Counter to the conventional practice, the equations are left in dimensional form and no special treatment has been given to the location of the upper bound of integration. We expect the integral relation to be applied as a post-processing step for scale-resolving data arising from, for example, high-resolution Particle Image Velocimetry or numerical simulations. We show the contributions of the various terms of the integral identities, the associated error in the estimate, and outline practical considerations when estimating the wall shear stress for complex flow conditions. 

This paper is organized as follows. In Section \ref{sec:Mathematical Formulation} the assumptions, data requirements, and mathematical formulation are listed. In Section \ref{sec:Demonstration Cases} eight demonstration cases are presented, followed by a discussion of the key insights and a summary of the results. Lastly, conclusions are summarized in Section \ref{sec:Conclusions}.

\section{Mathematical Formulation} \label{sec:Mathematical Formulation}

The integral formulation begins with the general Favre-averaged momentum equation:
\begin{equation}
    \pp{}{t}\left[\overline{\rho}\widetilde{u_i}\right] + \pp{}{x_j}\left[\overline{\rho}\widetilde{u_i}\widetilde{u_j}+\overline{p}\delta_{ij}+\overline{\rho u_i'' u_j''}-\overline{\tau_{ij}}\right] = 0 \label{eq:stream-momentum-eqn} 
\end{equation}
where $t$ denotes time, $\rho$ denotes density, $\overline{\;\cdot\;}$ denotes a Reynolds averaged quantity, $u_i$ denotes the velocity component in the $i$ spatial dimension using indicial notation, $\widetilde{\;\cdot\;}$ denotes a Favre averaged quantity, $x_i$ denotes the spatial coordinate in the $i$ direction, $p$ denotes static pressure, $\cdot''$ denotes a fluctuation about a Favre averaged quantity, and $\tau_{ij}$ denotes the viscous stress tensor. The viscous stress tensor is defined:
\begin{align}
\tau_{ij} & = 2\mu S_{ij} -\frac{2\mu}{3}\pp{u_k}{x_k}\delta_{ij} \label{eq:tau_ij}\\
    S_{ij} & = \frac{1}{2}\left(\pp{u_i}{x_j}+\pp{u_j}{x_i}\right) \label{eq:S_ij}
\end{align}
where $\mu$ denotes the dynamic viscosity, $S_{ij}$ denotes the strain rate tensor, and $\delta_{ij}$ denotes the Kronecker delta. Assuming a steady two-dimensional baseflow, and $\overline{\tau_{ij}''} \ll \widetilde{\tau_{ij}}$, the streamwise Favre-averaged momentum equation reads:
\begin{equation}\label{eq:stream-momentum-eqn-tau_expanded}
    \pp{}{x}\left[\overline{\rho}\widetilde{u}\widetilde{u}\right] + \pp{}{y}\left[\overline{\rho}\widetilde{u}\widetilde{v}\right]
    + \pp{\overline{p}}{x} + \pp{}{x}\left[\overline{\rho u'' u''}\right] + \pp{}{y}\left[\overline{\rho u'' v''}\right] - \pp{\overline{\tau_{11}}}{x} - \pp{\overline{\tau_{12}}}{y}= 0
\end{equation}
where now $x$ denotes the streamwise spatial coordinate, $y$ denotes the wall-normal spatial coordinate, $u$ denotes the streamwise velocity component, and $v$ denotes the wall-normal velocity component. The components of the viscous stress tensor which remain are:
\begin{align}
    \overline{\tau_{11}} &= \frac{4\overline{\mu}}{3}\pp{\widetilde{u}}{x} - \frac{2\overline{\mu}}{3}\pp{\widetilde{v}}{y} \label{eq:tau_11} \\
    \overline{\tau_{12}} &= \overline{\mu}\pp{\widetilde{u}}{y} + \overline{\mu}\pp{\widetilde{v}}{x} \label{eq:tau_12}    
\end{align}

The continuity equation:
\begin{equation}
\pp{\left[\overline{\rho}\widetilde{u}\right]}{x} + \pp{\left[\overline{\rho}\widetilde{v}\right]}{y} =0 \label{eq:continuity} 
\end{equation}
can be used to simplify the first two terms in Eq.~\ref{eq:stream-momentum-eqn-tau_expanded}. Further, recognizing $\overline{\rho}\widetilde{v}$ can be substituted from Eq.~\ref{eq:continuity-solve-rhov} as follows:
\begin{equation}
    \overline{\rho}\widetilde{v}= -\int\pp{\left[\overline{\rho}\widetilde{u}\right]}{x} dy     \label{eq:continuity-solve-rhov}
\end{equation}
The explicit dependence on wall-normal velocity can be eliminated from the streamwise momentum equation. This results in the following form of the streamwise momentum equation:

\begin{equation}
    \overline{\rho}\widetilde{u}\pp{\widetilde{u}}{x} - \pp{\widetilde{u}}{y} \int\pp{\left[\overline{\rho}\widetilde{u}\right]}{x}dy
    + \pp{\overline{p}}{x} + \pp{}{x}\left[\overline{\rho u'' u''}\right] + \pp{}{y}\left[\overline{\rho u'' v''}\right] - \pp{\overline{\tau_{11}}}{x} - \pp{\overline{\tau_{12}}}{y}= 0   
    \label{eq:stream-momentum-eqn-final}
\end{equation}

At this stage the governing momentum equation in the form presented in Eq. \ref{eq:stream-momentum-eqn-final} is a balance of forces per unit volume. If this equation is integrated in the wall-normal direction, the terms become a force per unit area, or a stress, and the equation becomes a stress balance at each wall-normal location. By integrating from the wall $y=0$ to some $y=y_P$ in the boundary layer, the wall-shear stress, defined as the lower bound of integration $\tau_w = \left.\tau_{12}\right|_{y=0}$, can be isolated from the equation and determined. The present formulation only integrates once in the wall-normal direction to eliminate the need for data throughout the entire boundary layer. Moreover, it avoids the spurious dependence on the upper boundary location as highlighted by Xu et al. \cite{Xu23} when doing twofold or threefold integrations. The outcome of integrating once in the wall normal direction is as follows:

\begin{align}
    \int_0^y\overline{\rho}\widetilde{u}&\pp{\widetilde{u}}{x}dy - \int_0^y\left(\pp{\widetilde{u}}{y} \int\pp{\left[\overline{\rho}\widetilde{u}\right]}{x}dy\right)dy
    + \int_0^y\pp{\overline{p}}{x}dy + \nonumber\\ &\int_0^y\pp{}{x}\left[\overline{\rho u'' u''}\right]dy + \left[\overline{\rho u'' v''}\right]_0^y - \int_0^y\pp{\overline{\tau_{11}}}{x}dy- \left.\overline{\tau_{12}}\right|_0^y = 0
    \label{eq:wall-normal-integral-step1}
\end{align}

The integral relation, Eq. \ref{eq:wall-shear-stress-balance-terms-annotated}, is found after the nested integral in second term from the left of Eq. \ref{eq:wall-normal-integral-step1} is expanded using integration by parts. In addition to recognizing the following bounds of integration:
\begin{align}
    \left.\widetilde{u}\right|_0 &= 0 \;\;\;\;\; \text{because of the no-slip condition}\nonumber\\
    \left.\widetilde{u}\right|_y &= 0 \;\;\;\;\; \text{is the Favre-averaged streamwise velocity at a given $y$ location}\nonumber\\
    \left.\overline{\rho u''v''}\right|_{0} &= 0 \;\;\;\;\; \text{because } u''=v''=0 \;\; \text{at } y=0 \nonumber \\
    \left.\overline{\rho u''v''}\right|_{y} & \;\;\;\;\;\;\;\;\;\;\;\; \text{is simply the value of the given term at } y \nonumber \\
    \left.\overline{\tau_{12}}\right|_0 & = \tau_w \;\;\; \text{wall shear stress of interest} \nonumber \\
    \left.\overline{\tau_{12}}\right|_{y} & \;\;\;\;\;\;\;\;\;\;\;\; \text{is simply the value of the given term at } y \nonumber
\end{align}

\begin{align}
    \tau_w = \underbrace{\overline{\tau_{12}}}_{\text{I}} - \underbrace{\overline{\rho u'' v''}}_{\text{II}} -
    \underbrace{\int_0^y\pp{\overline{p}}{x}dy}_{\text{III}} + \;\underbrace{\widetilde{u}\int_0^y\pp{\left[\overline{\rho}\widetilde{u}\right]}{x}dy}_{\text{IV}} \; &- \underbrace{\int_0^y\overline{\rho}\widetilde{u} \pp{\widetilde{u}}{x}dy}_{\text{V}} \;- \underbrace{\int_0^y\widetilde{u}\pp{\left[\overline{\rho}\widetilde{u}\right]}{x}dy}_{\text{VI}} \nonumber \\ & + \underbrace{\int_0^y\pp{\overline{\tau_{11}}}{x}dy}_{\text{VII}} \; -
    \underbrace{\int_0^y\pp{}{x}\left[\overline{\rho u'' u''}\right]dy}_{\text{VIII}} 
    \label{eq:wall-shear-stress-balance-terms-annotated}
\end{align}

Equation \ref{eq:wall-shear-stress-balance-terms-annotated} has been formulated to remove all dependence on mean wall-normal velocity; the tradeoff being that at least two wall-normal boundary layer profiles with sufficient resolution to compute wall-normal and streamwise gradients are required. This form is presented here in the case that the only available data is limited to wall-parallel velocity with sufficient streamwise information. Section \ref{sec:Eliminating-Streamwise-Gradients} includes discussion regarding the elimination of the potentially troublesome streamwise gradients.

\subsection{Description of Terms}
Term I represents the viscous shear stress and term II represents the Reynolds (turbulent) shear stress. Terms I and II are often combined into a total shear stress $\overline{\tau} = \overline{\tau_{12}}-\overline{\rho u''v''}$. Term III is the pressure gradient term and is important for nonzero pressure gradient flows. Terms IV, V, and VI all result from the $\pp{\widetilde{u}}{y}\int\pp{\left[\overline{\rho}\widetilde{u}\right]}{x}dy$ term from Eq. \ref{eq:stream-momentum-eqn-final} and become important outside of the viscous sublayer, or approximately $y/\delta>0.1$. Term VII comes from the integration of the viscous normal stress and term VIII arises from integrating the streamwise Reynolds (turbulent) normal stress. Analogous to terms I and II which are for the shear stress component of the stress tensor, terms VII and VIII are the viscous and turbulent counterparts for the normal component of the stress tensor. The $\tau_{11}$ in term VII is related to the dilatation (compression) and is practically zero everywhere except in regions of high compression like a shock, so in the wall-bounded flows of interest it is often neglected. Term VIII is also found to be small in the boundary layer and can be neglected. Terms involving wall-normal integrals are sensitive to the lower boundary condition in so much as the lower bound of integration will affect the outcome of the subsequent integral regardless of the upper bound of integration. Being a simple stress balance, the terms will be valid and may be integrated up to any $y$-location above the wall at $y=0$.

\subsection{Eliminating Streamwise Gradients}\label{sec:Eliminating-Streamwise-Gradients}
In practice, terms IV, V, and VI present challenges because of the presence of streamwise gradients. Ideally, all derivatives are with respect to the wall normal coordinate, and so only one wall-normal profile is needed for the stress balance in Eq. \ref{eq:wall-shear-stress-balance-terms-annotated}. The streamwise gradients in terms IV and VI can be replaced with wall-normal gradients by making use of the continuity equation. 

\begin{equation}
    \pp{\left[\overline{\rho}\widetilde{u}\right]}{x} = - \pp{\left[\overline{\rho}\widetilde{v}\right]}{y}
    \label{eq:continuity-gradient-swap}
\end{equation}

Similarly for term V, the $\pp{\widetilde{u}}{x}$ gradient can be replaced with wall-normal derivatives only (except the pressure gradient term) by manipulating the steady, two-dimensional momentum equation. Consistent with the previous discussion, the dilatation and normal component of the Reynolds stress tensor are neglected.
\begin{equation}
    \pp{\widetilde{u}}{x} = \frac{1}{\overline{\rho}\widetilde{u}}\left(-\overline{\rho}\widetilde{v}\pp{\widetilde{u}}{y} 
    - \pp{\overline{p}}{x} - \pp{}{y}\left[\overline{\rho u'' v''}\right] + \pp{\overline{\tau_{12}}}{y}\right)
    \label{eq:momentum-gradient-swap}
\end{equation}

Substituting the expressions from Eqs. \ref{eq:continuity-gradient-swap} and \ref{eq:momentum-gradient-swap} into the integral equation Eq. \ref{eq:wall-shear-stress-balance-terms-annotated}, integrating, and simplifying does not provide additional benefit because terms end up canceling. Rather, the verbose form of the streamwise gradients should be pre-computed from Eqs. \ref{eq:continuity-gradient-swap} and \ref{eq:momentum-gradient-swap} respectively, and then subsequently used in  Eq. \ref{eq:wall-shear-stress-balance-terms-annotated}. The advantage being that now the entire stress balance only depends on terms with wall-normal gradients; therefore, only one wall-normal profile is needed. 

The final streamwise gradient to consider is the pressure gradient, $\pp{\overline{p}}{x}$. Fortunately, this term proves to be the least sensitive of all the streamwise gradient terms. By making use of the constant wall-normal pressure boundary layer assumption, it can be found from boundary layer edge or surface conditions, either directly from the data at the edge or surface, or by some reduced order method like modified Newtonian theory \cite{Lees1955}.

For the present work, all gradients are computed from a first order forward or backward difference, unless otherwise specified. In cases with limited wall-normal resolution, a spline can be interpolated to the wall-normal profiles to improve the estimates; however, this is seldom needed. Lastly, when implementing the integral relation, the boundary layer profiles should be transformed to local coordinates parallel and normal to the wall. 

\section{Demonstration Cases} \label{sec:Demonstration Cases}
Eight test cases have been selected to demonstrate the importance of each term in Eq. \ref{eq:wall-shear-stress-balance-terms-annotated} and the overall ability to estimate the wall shear stress. The conditions selected are high-speed, compressible flows and effort was made to find complex flows involving pressure gradients, surface roughness, and surface curvature. Table \ref{tb:summary-of-cases} summarizes the eight test cases selected, and a check mark identifies which feature of interest is present. The possible features of interest are compressibility (Comp.), turbulence (Turb.), pressure gradients, surface curvature, and surface roughness.

\begin{table}[h!]
\newcolumntype{P}[1]{>{\centering\arraybackslash}p{#1}}
\centering
    \caption{Summary of demonstration cases.}
    \begin{tabular}{c*{5}{P{2cm}}}
    \hline
    Case & Comp. & Turb. & \makecell[c]{Pressure\\[-10pt] Gradients} & \makecell[c]{Surface\\[-10pt] Curvature} & \makecell[c]{Surface\\[-10pt] Roughness} \\ \hline\hline
    \makecell[l]{Mach 2.5, ZPG, Laminar,\\ Flat Plate [Present Work]} & \checkmark & & & & \\\hline
    \makecell[l]{Mach 2.5, Smooth-Wall,\\Turbulent Channel Flow \cite{Gerolymos23,Gerolymos24,Gerolymos_database24}} & \checkmark & \checkmark & \checkmark & & \\\hline
    \makecell[l]{Mach 4.0, Rough-Wall,\\Turbulent Channel Flow \cite{Modesti22,Modesti22_database}} & \checkmark & \checkmark & \checkmark & & \checkmark \\\hline
    \makecell[l]{Mach 2.9, ZPG, Rough-Wall,\\Turbulent Flat Plate [Present Work]} & \checkmark & \checkmark & &  & \checkmark \\\hline
    \makecell[l]{Mach 6.0, Laminar, Hypersonic,\\Blunt Body [Present Work]} & \checkmark & & \checkmark & \checkmark & \\\hline
    \makecell[l]{Mach 4.9, Smooth, ZPG Flat\\Plate, Turbulent Flow \cite{NASA-LARC-Turb,nicholson24}} & \checkmark & \checkmark & \checkmark & & \\\hline
    \makecell[l]{Mach 4.9, Smooth, Forward-Facing\\Wall, Turbulent Flow \cite{NASA-LARC-Turb,nicholson24}} & \checkmark & \checkmark & \checkmark & \checkmark & \\\hline
    \makecell[l]{Mach 4.9, Smooth, Backward-Facing\\Wall, Turbulent Flow \cite{NASA-LARC-Turb,nicholson24}} & \checkmark & \checkmark & \checkmark & \checkmark & \\ \hline
    \end{tabular}
    \label{tb:summary-of-cases}
\end{table}

\subsection{Mach 2.5, ZPG, Laminar Flat Plate}\label{sec:Mach 2.5, ZPG, Laminar Flat Plate}
The first test case is a zero-pressure-gradient (ZPG), laminar, smooth-wall, flat plate boundary layer. The conditions for the test are inspired by the turbulent DNS database from Zhang et al. \cite{Zhang18}; specifically, the M2p5 case with freestream conditions $M_e=2.50$, $U_e=823.6$ [m/s], $\rho_e=0.100$ [kg/m$^3$], $T_e= 270.0$ [K], and an isothermal wall $T_w=568.0$ [K]. Solving the compressible, ZPG similarity solution of the form of Eq. \ref{eq:momt-fs-ss} and \ref{eq:nrg-fs-ss} yields wall-normal profiles for the given conditions. 
\begin{align}
    \left(Cf''\right)'+ff''&=\beta\left(f'^2-g\right)\label{eq:momt-fs-ss}\\
    \left(Cg'\right)'+Prfg'&=-PrC(\gamma-1)M_e^2f''^2\label{eq:nrg-fs-ss}
\end{align}
Similarity variables $f$ and $g$ are defined as follows:
\begin{equation}
    f'=\frac{u}{U_e} \;\;\;\;\;\;\;\;\;\; \text{and} \;\;\;\;\;\;\;\;\;\; g=\frac{T}{T_e}
\end{equation}
Differentiation with respect to $\eta$ is denoted by $\cdot'$ where the Illingworth transformation~\cite{Illingworth50} is used for streamwise and wall-normal similarity coordinates $\xi$ and $\eta$ as follows:
\begin{equation}
    \xi(x) = \int_0^x\rho_e(x)U_e(x)\mu_e(x)dx \;\;\;\;\;\;\;\;\;\; \text{and} \;\;\;\;\;\;\;\;\;\; \eta(x,y) = \frac{U_e}{\sqrt{2\xi}}\int_0^y\rho dy
    \label{eq:illingworth-transformation}
\end{equation}
The parameters $\beta$ and $C$ are the pressure gradient parameter and the Chapman-Rubesin parameter respectively defined as follows:
\begin{equation}
\beta = \frac{2\xi}{M_e}\frac{dM_e}{d\xi}
\label{eq:beta-pressure-gradient}
\end{equation}
\begin{equation}
    C = \frac{\rho\mu}{\rho_e\mu_e}
    \label{eq:chapman-rubesin}
\end{equation}
Thermodynamic properties are related using the ideal gas equation of state and $Pr$ denotes the Prandtl number which is set to $Pr = 0.71$ for the case presented.

\begin{figure}[h!]
\centering
\captionsetup{justification=centering}
\begin{subfigure}{0.5\textwidth}
    \centering
    \captionsetup{justification=centering}
    \includegraphics[width=\textwidth]{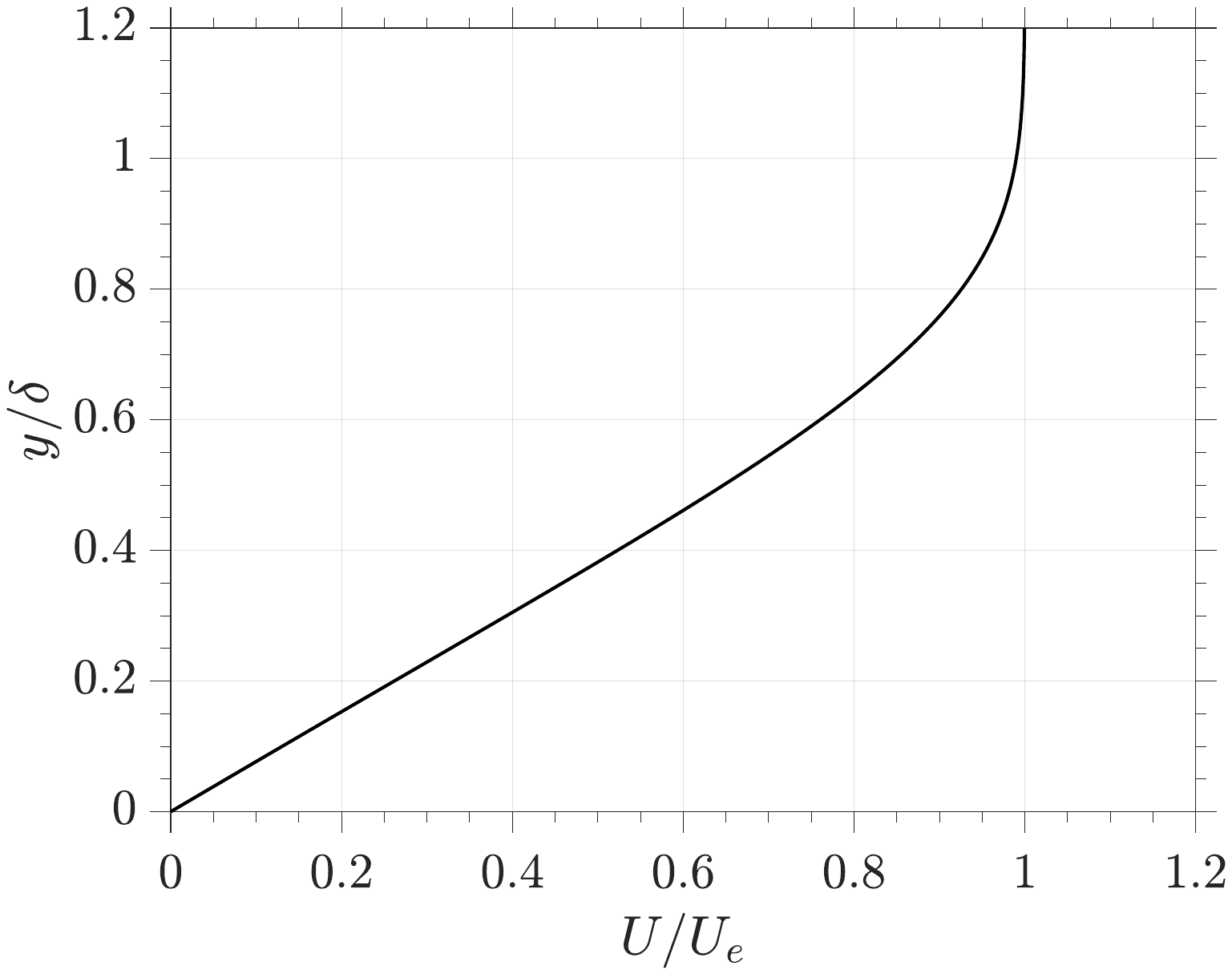}
    \caption{Wall-parallel velocity}
    \label{fig:Blasius-BL-upara}
\end{subfigure}%
~
\begin{subfigure}{0.5\textwidth}
    \centering
    \captionsetup{justification=centering}
    \includegraphics[width=\textwidth]{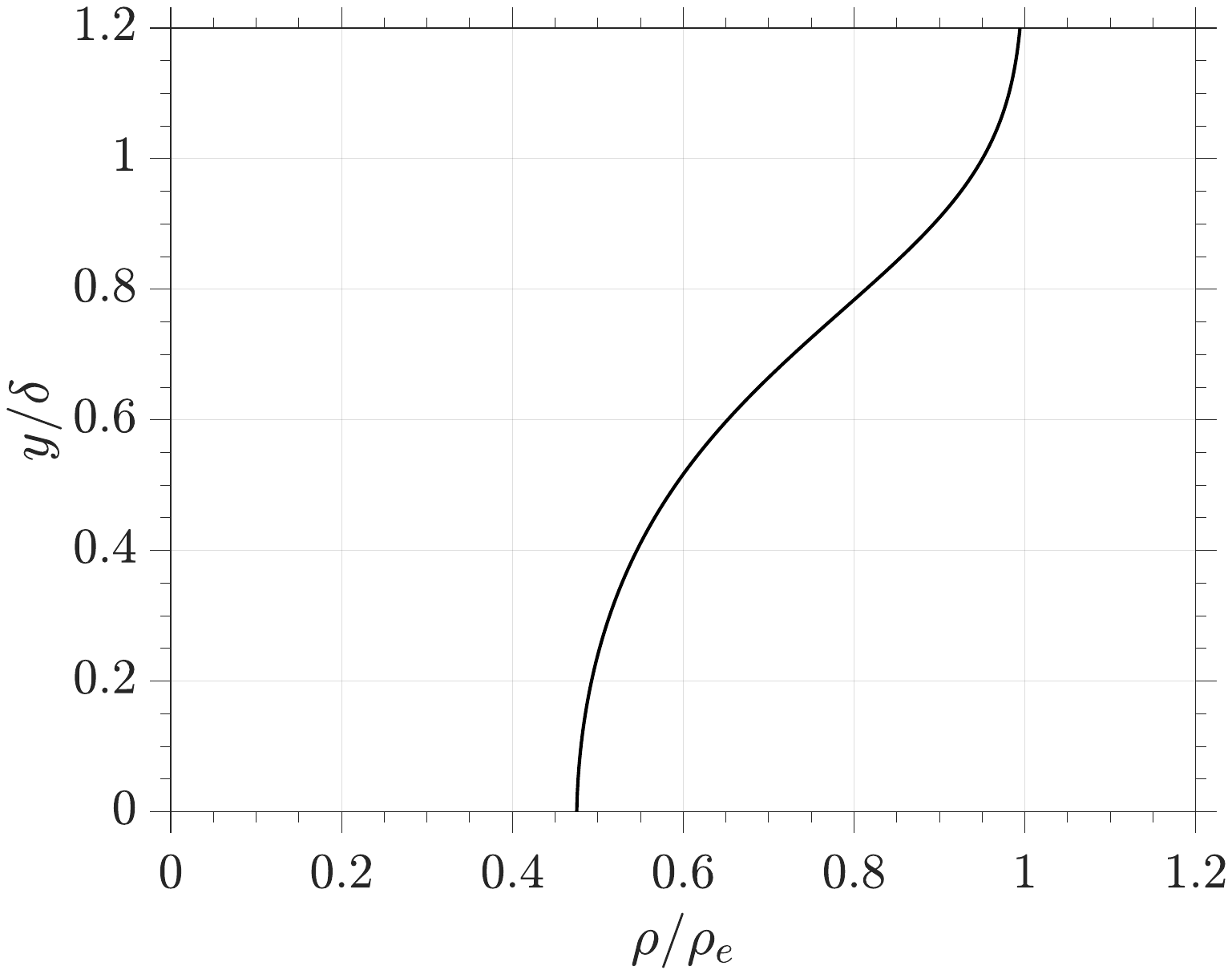}
    \caption{Density}
    \label{fig:Blasius-BL-rho}
\end{subfigure}
\caption{Compressible, ZPG, flat plate boundary layer wall-normal profiles.}
\label{fig:Blasius-BL-Profiles}
\end{figure}

For the ZPG flat plate, the pressure gradient parameter $\beta=0$. For this example case, the similarity solution is solved using the shooting method. The boundary layer velocity and density profiles for the selected conditions are shown in Fig. \ref{fig:Blasius-BL-Profiles}. In this case, it is easy to obtain two wall-normal profiles at different streamwise locations; therefore Eq. \ref{eq:wall-shear-stress-balance-terms-annotated} can be used directly without substitution of the streamwise gradients. Two wall-normal profiles were taken at streamwise locations of $x=0.014846$ [m] and $x=0.014900$ [m] and a simple forward difference is used to compute the gradients.

\begin{figure}[h!]
\centering
\captionsetup{justification=centering}
\begin{subfigure}{0.5\textwidth}
    \centering
    \captionsetup{justification=centering}
    \includegraphics[width=\textwidth]{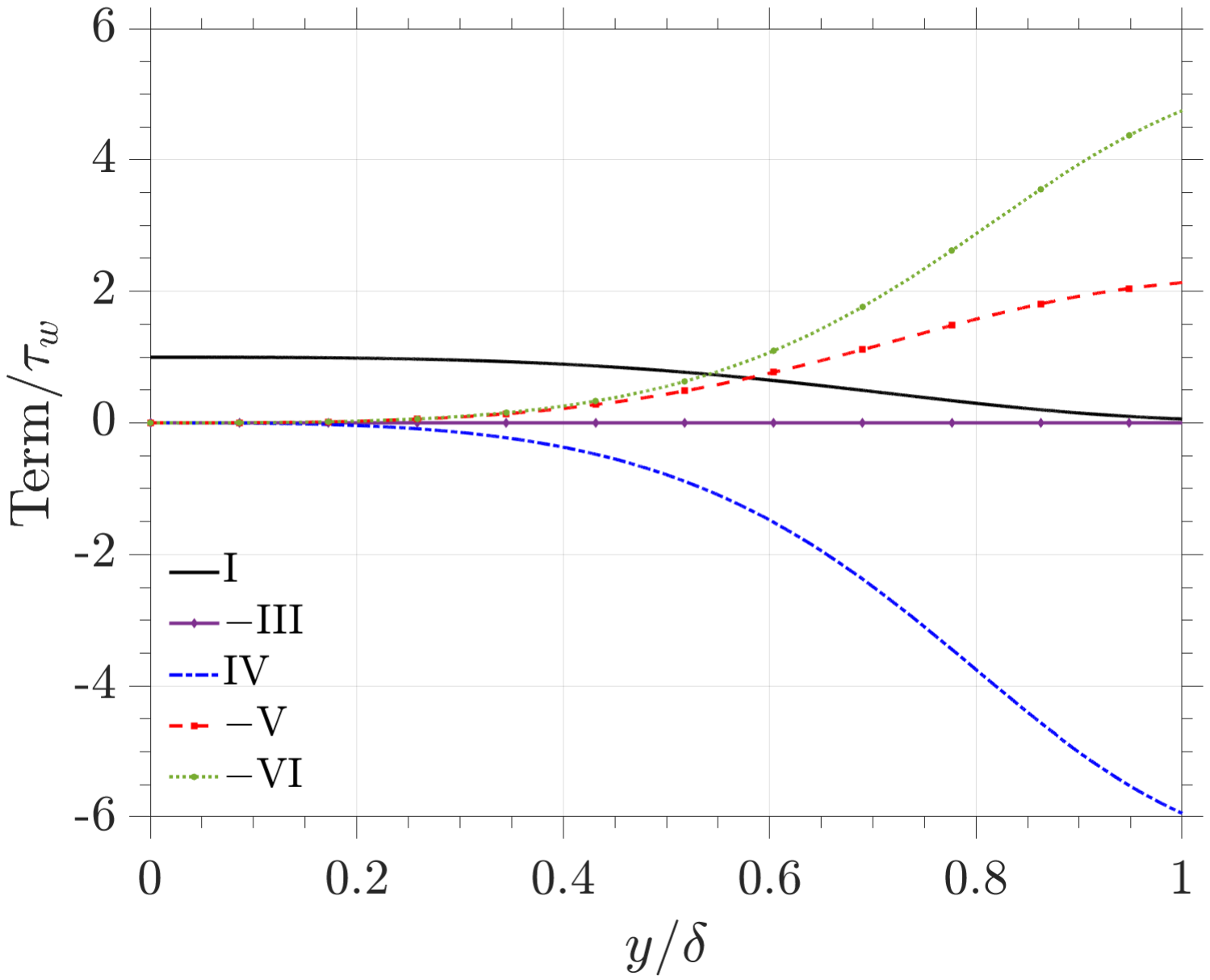}
    \caption{Contributing terms}
    \label{fig:Blasius-BL-terms}
\end{subfigure}%
~
\begin{subfigure}{0.5\textwidth}
    \centering
    \captionsetup{justification=centering}
    \includegraphics[width=\textwidth]{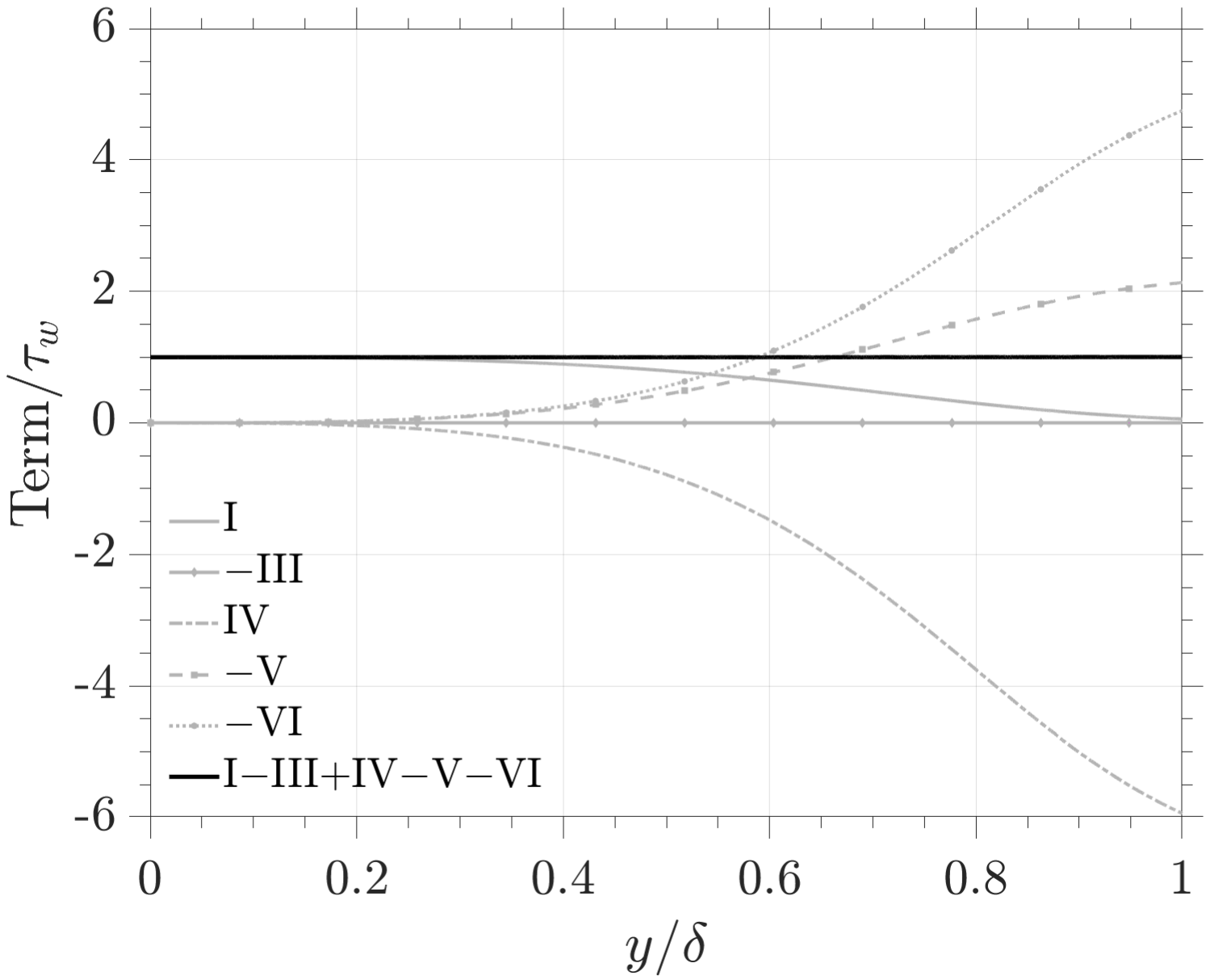}
    \caption{Sum of terms}
    \label{fig:Blasius-BL-termsSUM}
\end{subfigure}
\caption{ZPG flat plate boundary layer contribution of each stress term as a function of wall-normal distance. Vertical sum results in constant wall shear stress estimate.}
\label{fig:Blasius-BL-TermsSum}
\end{figure}

For the laminar ZPG boundary layer only terms I, III, IV, V, and VI are shown because term II is zero for laminar flow and terms VII and VIII can be neglected. Figure \ref{fig:Blasius-BL-terms} shows the value of each term in Eq. \ref{eq:wall-shear-stress-balance-terms-annotated} as a function of wall-normal position, normalized by boundary layer height $\delta$. Figure \ref{fig:Blasius-BL-termsSUM} has been included to illustrate that summing the value of each term at a given wall-normal position ($y/\delta$) results in the value of the wall shear stress. If the boundary layer profiles provided are sufficiently accurate, the value of the wall shear stress sum will be constant across the entire boundary layer height. This constant sum of all the terms for the entire boundary layer should occur regardless of the flow conditions such as turbulence, pressure gradient, or roughness (provided the assumptions in the derivation of Eq. \ref{eq:wall-shear-stress-balance-terms-annotated} are satisfied). For subsequent test cases, only plots of the type of Fig. \ref{fig:Blasius-BL-terms} will be shown and the summation like Fig. \ref{fig:Blasius-BL-termsSUM} will be omitted. Given that the present method of calculating the wall shear stress provides a constant estimate across the entire boundary layer height, the wall shear stress can be estimated either by taking the average of the computed sum across the boundary layer, or simply selecting the sum at a given wall-normal location. For example, in situations with surface roughness, it may be advantageous to select a location further from the wall to avoid corruption of the near-wall flow due to the presence of the roughness.

The shear stress computed from the velocity gradient at the wall is $\tau_w = \left.\mu\frac{dU}{dy}\right|_w=77.96$ [Pa]. The average value from $y/\delta=0$ to $y/\delta=1$ of the integral method estimate is $\tau_w = 78.02$ [Pa]. For practical purposes, the two results are exact, with the small percent error of 0.08\% due to numerical errors associated with the approximation of the gradients using a finite difference technique.

\subsection{Mach 2.5, Smooth-Wall, Turbulent Channel Flow}
\begin{figure}[h!]
\centering
\captionsetup{justification=centering}
\begin{subfigure}{0.5\textwidth}
    \centering
    \captionsetup{justification=centering}
    \includegraphics[width=\textwidth]{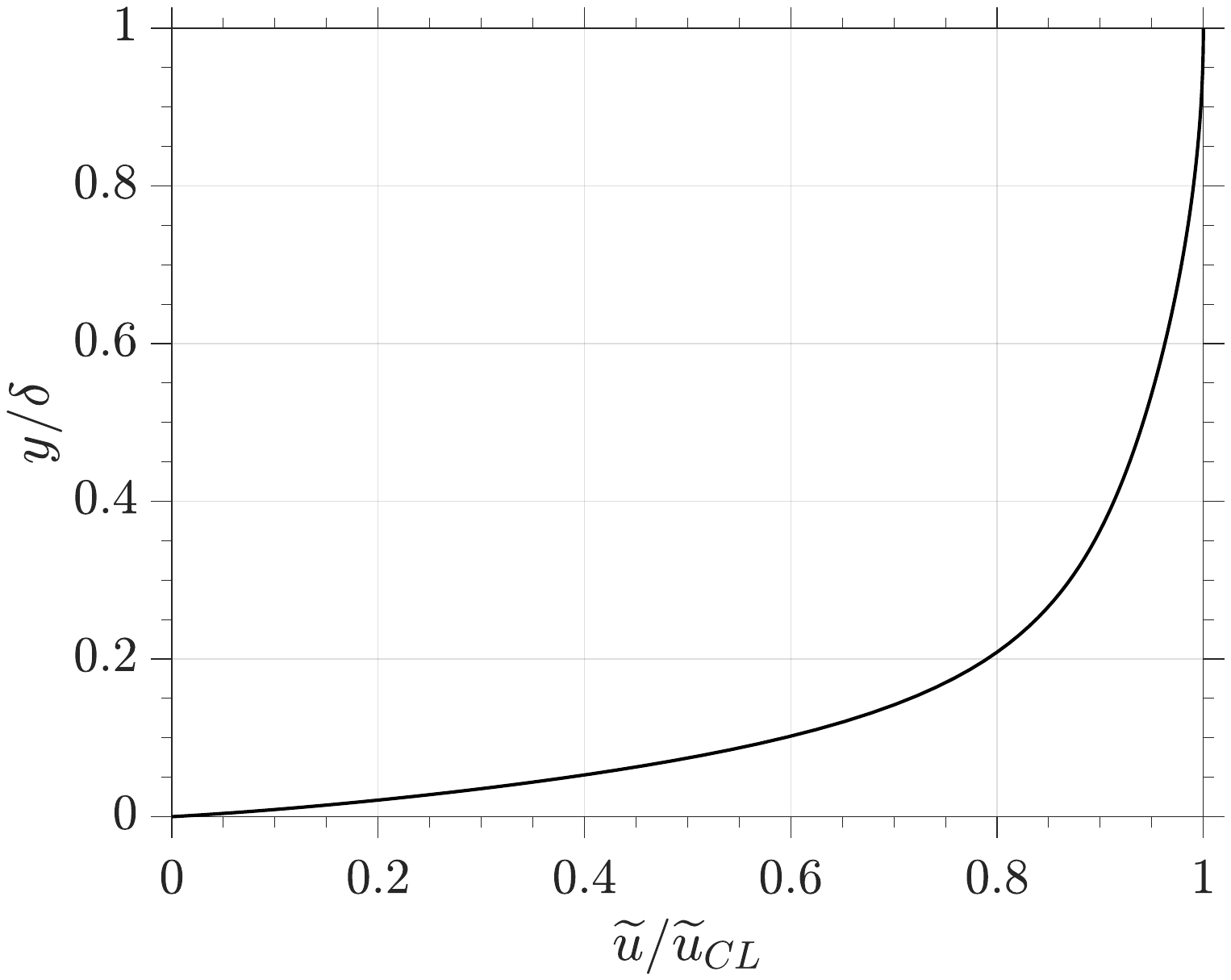}
    \caption{Streamwise velocity}
    \label{fig:comp-smooth-turb-channel-U}
\end{subfigure}%
~
\begin{subfigure}{0.5\textwidth}
    \centering
    \captionsetup{justification=centering}
    \includegraphics[width=\textwidth]{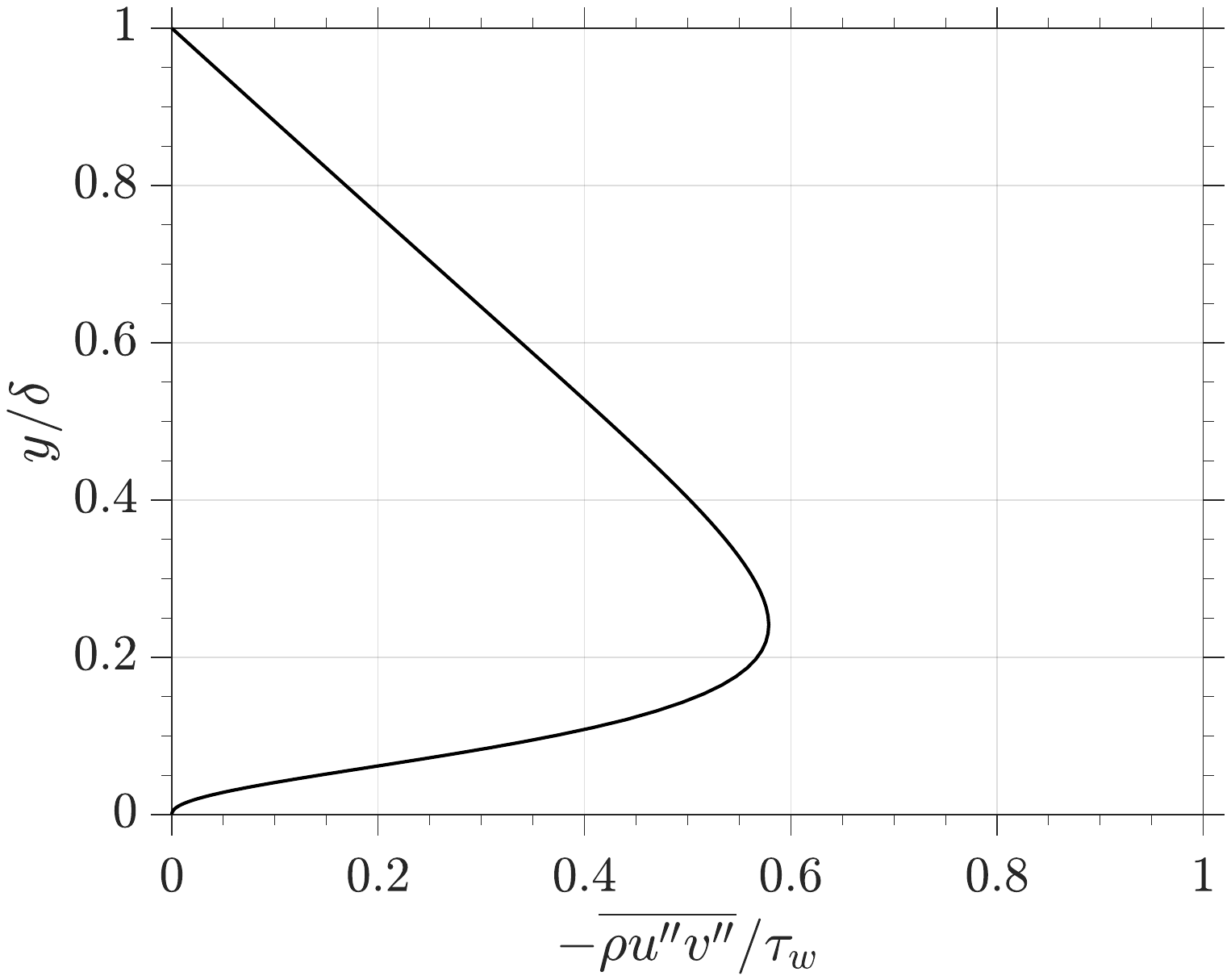}
    \caption{Reynolds shear stress}
    \label{fig:comp-smooth-turb-channel-RS}
\end{subfigure}
\caption{Compressible, smooth-wall, turbulent channel flow wall-normal profiles.}
\label{fig:comp-smooth-turb-channel-Profiles}
\end{figure}

The next demonstration case is compressible, smooth-wall turbulent plane channel flow. Direct numerical simulation (DNS) data has been obtained from the database by Gerolymos and Vallet \cite{Gerolymos_database24}. The original physical analysis of the data by Gerolymos and Vallet is found in \cite{Gerolymos23,Gerolymos24}. The database contains 25 flow conditions determined by the corresponding Reynolds and Mach numbers. For the present work the $\overline{M}_{CL}=2.49$, $Re_{\tau*}=113$, perfect gas air, and isothermal wall $T_w=298$ [K] conditions are selected. The streamwise centerline Mach number is defined as $\overline{M}_{CL}=\overline{\left(\frac{u_{CL}}{a_{CL}}\right)}$ where $u_{CL}$ is the streamwise centerline velocity and $a_{CL}$ is the centerline speed of sound. The friction Reynolds number is defined as the Huang-Coleman-Bradshaw (HCB) friction Reynolds number $Re_{\tau*} = \frac{\overline{\rho}_{CL}}{\overline{\mu}_{CL}}\sqrt{\frac{\overline{\tau_w}}{\overline{\rho}_{CL}}}\delta$, where $\delta$ is the wall-to-centerline distance. The Favre-averaged streamwise velocity profile is shown in Fig. \ref{fig:comp-smooth-turb-channel-U} and the Reynolds shear stress profile is shown in Fig. \ref{fig:comp-smooth-turb-channel-RS}. Normalization of the streamwise-velocity is by the Favre-averaged centerline streamwise velocity and the Reynolds stress is normalized by the wall shear stress. Note that the $\tau_w$ from the nondimensionalization needs to be averaged because the flow is turbulent; however, for simplicity of notation the overline in $\overline{\tau_w}$ has been dropped to be consistent with the derivation presented in earlier sections.

\begin{figure}[h!]
\centering
\captionsetup{justification=centering}
\begin{subfigure}{0.5\textwidth}
    \centering
    \captionsetup{justification=centering}
    \includegraphics[width=\textwidth]{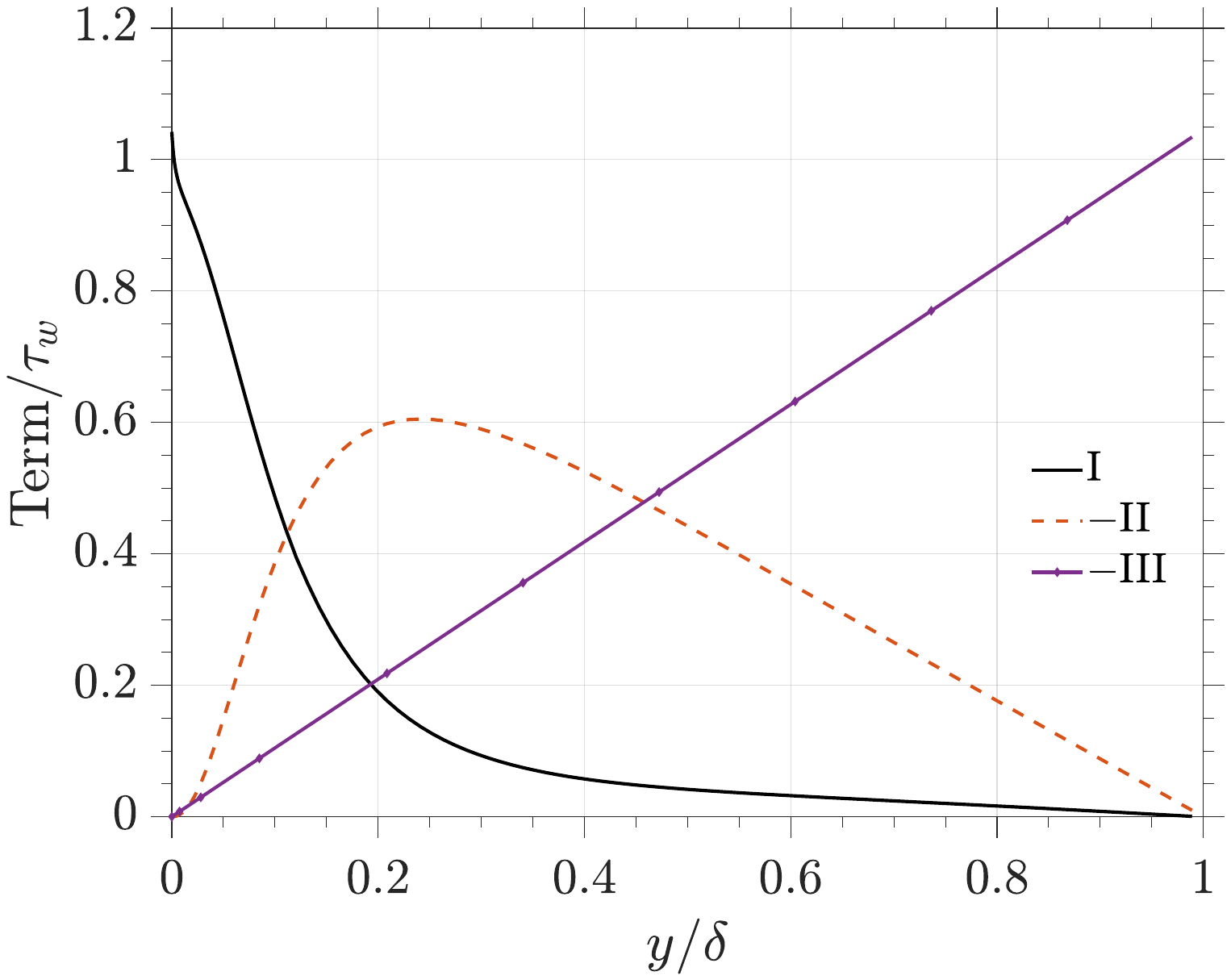}
    \caption{Outer scaling}
    \label{fig:comp-smooth-turb-channel-terms-outer}
\end{subfigure}%
~
\begin{subfigure}{0.5\textwidth}
    \centering
    \captionsetup{justification=centering}
    \includegraphics[width=\textwidth]{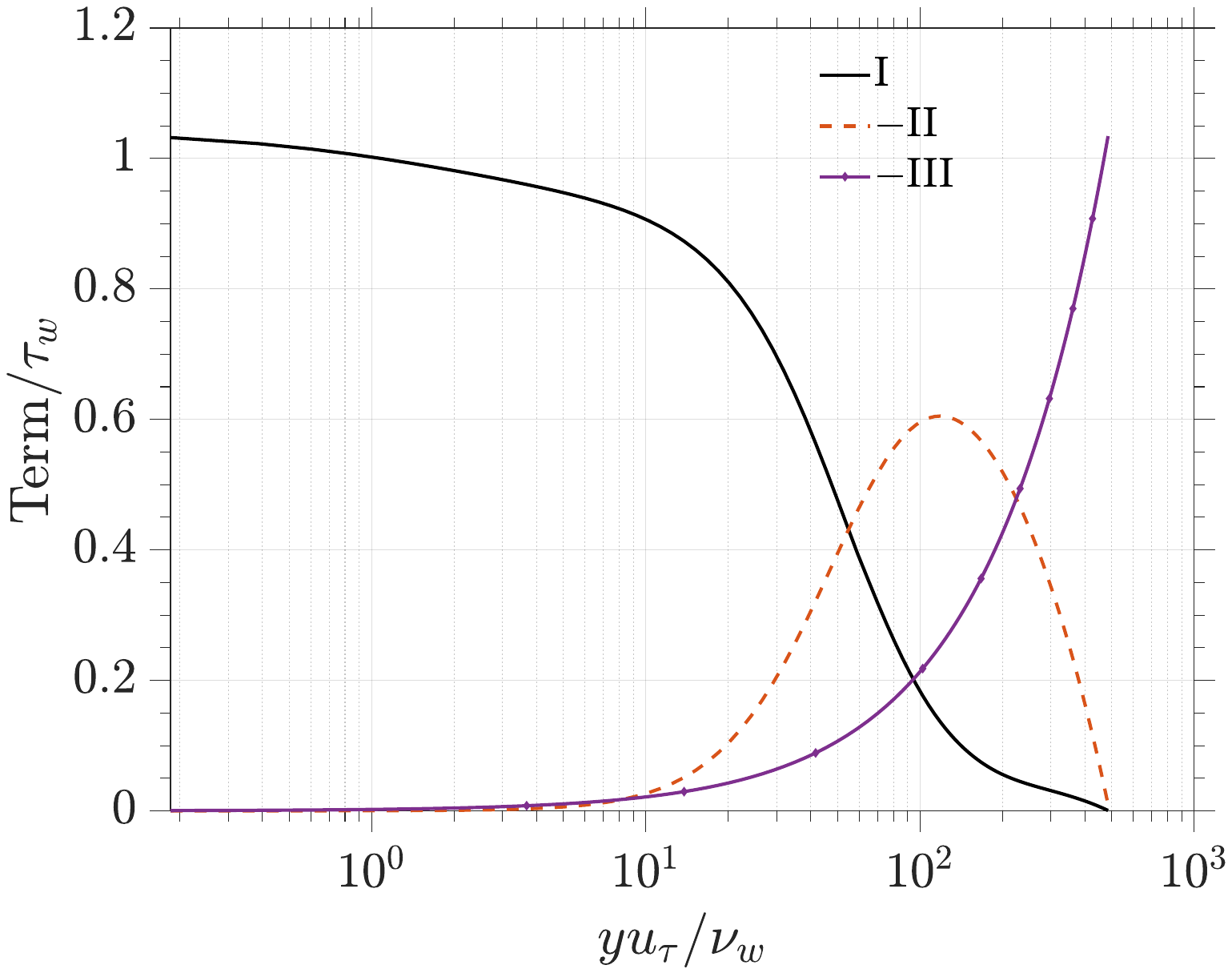}
    \caption{Inner scaling}
    \label{fig:comp-smooth-turb-channel-terms-plus}
\end{subfigure}
\caption{Compressible, smooth-wall, turbulent channel flow contribution of each stress term as a function of wall-normal distance.}
\label{fig:comp-smooth-turb-channel-terms}
\end{figure}

For fully-developed channel flow all streamwise gradients ($\pp{}{x}$ terms) will be zero except for $\pp{p}{x}$ which will be constant. Therefore, the effectiveness of the present method only needs one streamwise location's wall-normal profile and the overall pressure gradient. The only non-zero terms that contribute to the shear stress estimate are terms I, II, and III. The value of each term as a function of wall-normal distance are shown in Fig. \ref{fig:comp-smooth-turb-channel-terms}.

\begin{figure}[h!]
\centering
\captionsetup{justification=centering}
\begin{subfigure}{0.5\textwidth}
    \centering
    \captionsetup{justification=centering}
    \includegraphics[width=\textwidth]{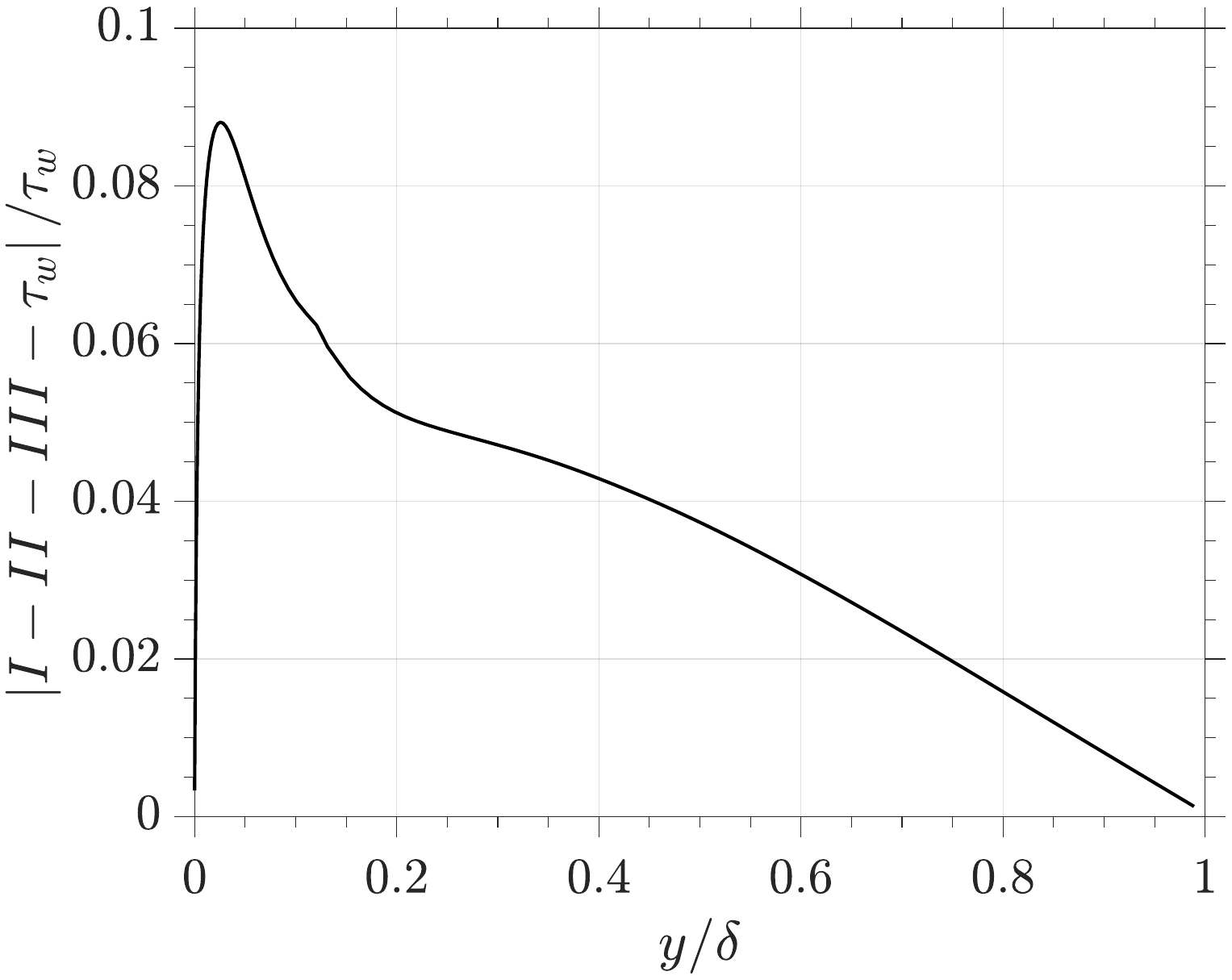}
    \caption{Outer scaling}
    \label{fig:comp-smooth-turb-channel-err-outer}
\end{subfigure}%
~
\begin{subfigure}{0.5\textwidth}
    \centering
    \captionsetup{justification=centering}
    \includegraphics[width=\textwidth]{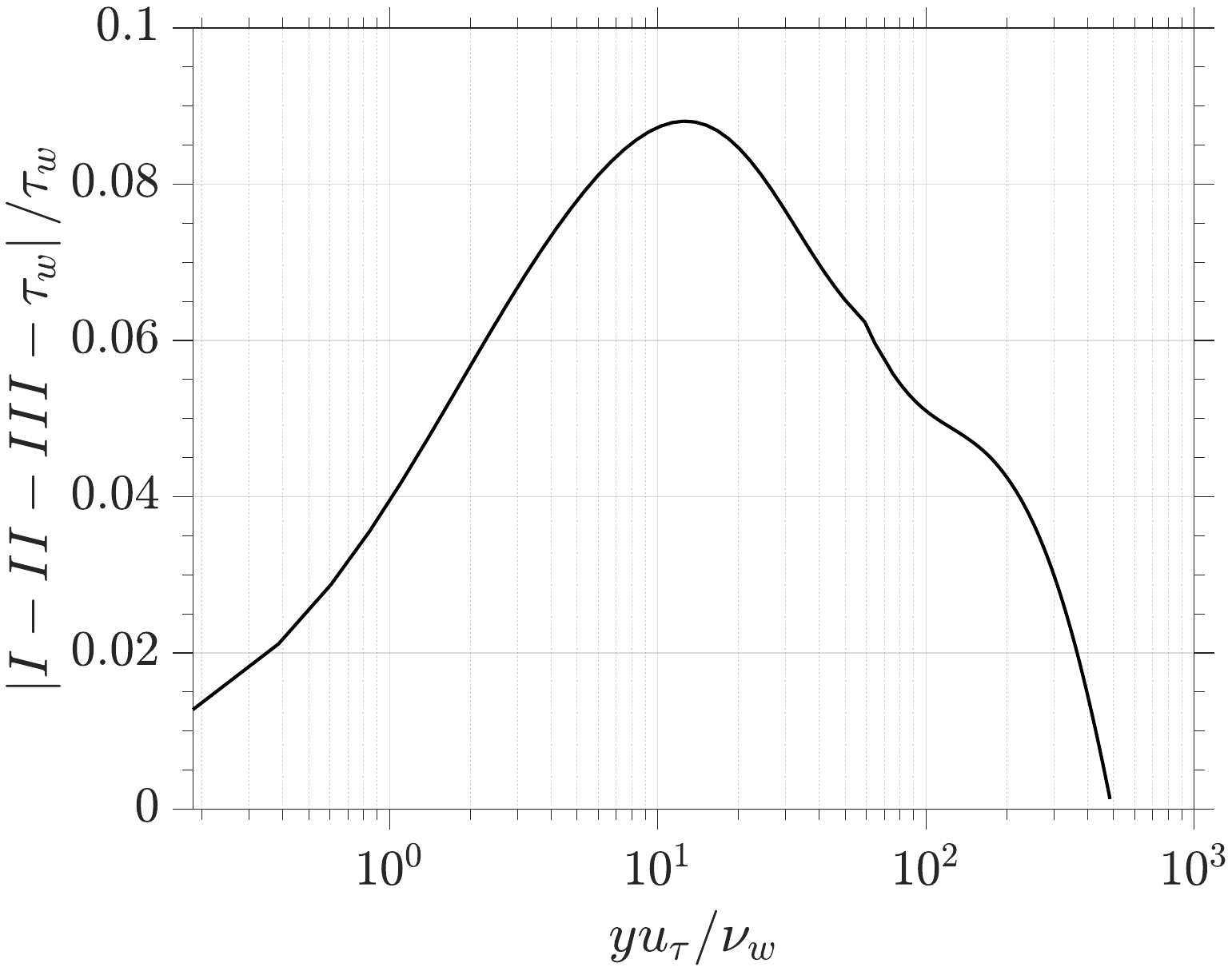}
    \caption{Inner scaling}
    \label{fig:comp-smooth-turb-channel-err-plus}
\end{subfigure}
\caption{Compressible, smooth-wall, turbulent channel flow error in shear stress estimate as a function of wall-normal distance.}
\label{fig:comp-smooth-turb-channel-err}
\end{figure}

The wall shear stress reported by the authors of the database is $\tau_w=269.73$ [Pa] \cite{Gerolymos_database24}, from the present stress balance the average is $\tau_w=257.91$ [Pa]. The percent error as a function of wall-normal distance between the present estimate and the reported value is shown in Fig. \ref{fig:comp-smooth-turb-channel-err}. The maximum percent error is 8.81\% with the average percent error at -4.38\%. The maximum error occurs in the buffer layer and then reduces to approximately zero at the channel centerline.

\subsection{Mach 4.0, Rough-Wall, Turbulent Channel Flow}
\begin{figure}[h!]
\centering
\captionsetup{justification=centering}
\begin{subfigure}{0.5\textwidth}
    \centering
    \captionsetup{justification=centering}
    \includegraphics[width=\textwidth]{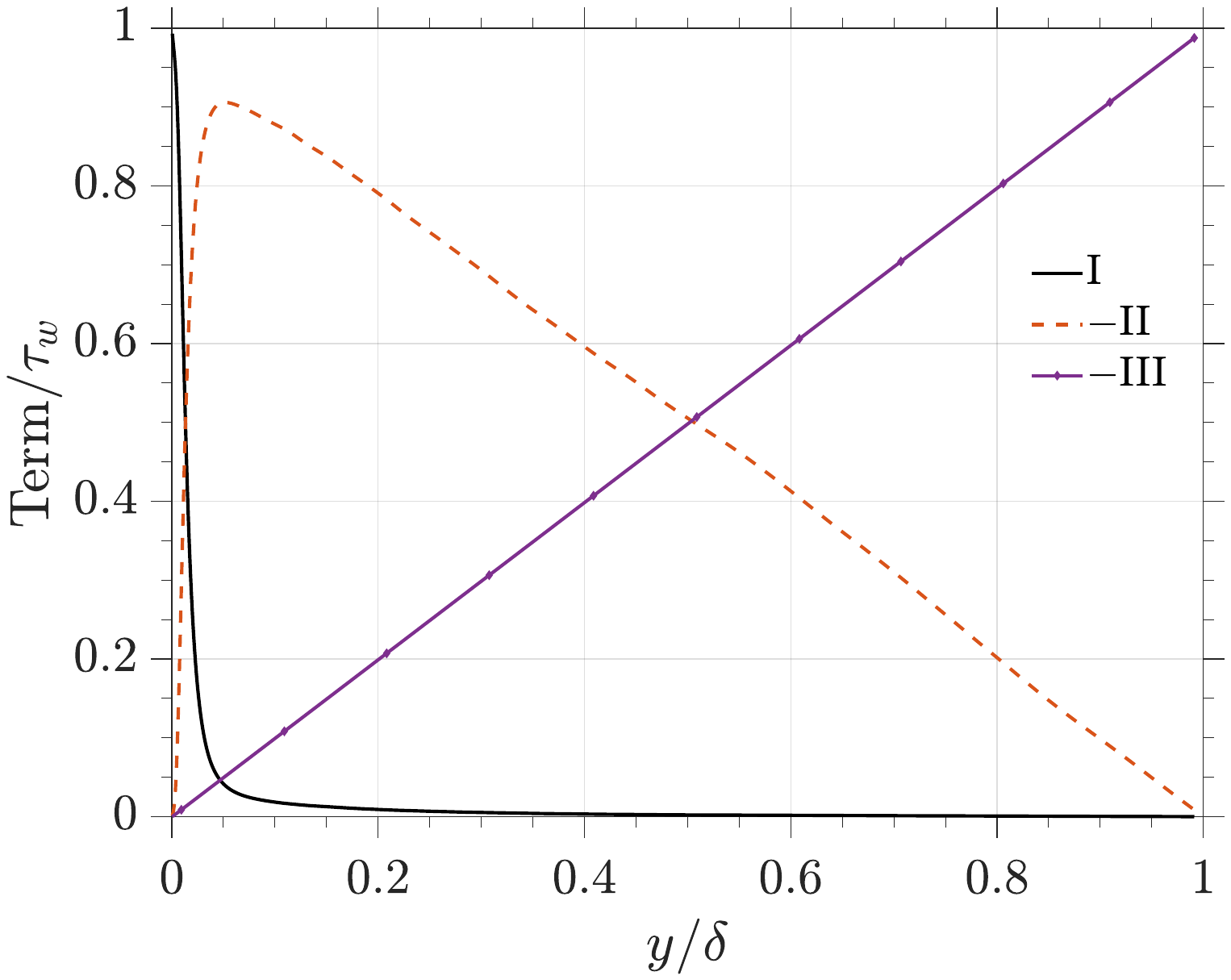}
    \caption{}
    \label{fig:comp-turb-channel-terms-smooth}
\end{subfigure}%
~
\begin{subfigure}{0.5\textwidth}
    \centering
    \captionsetup{justification=centering}
    \includegraphics[width=\textwidth]{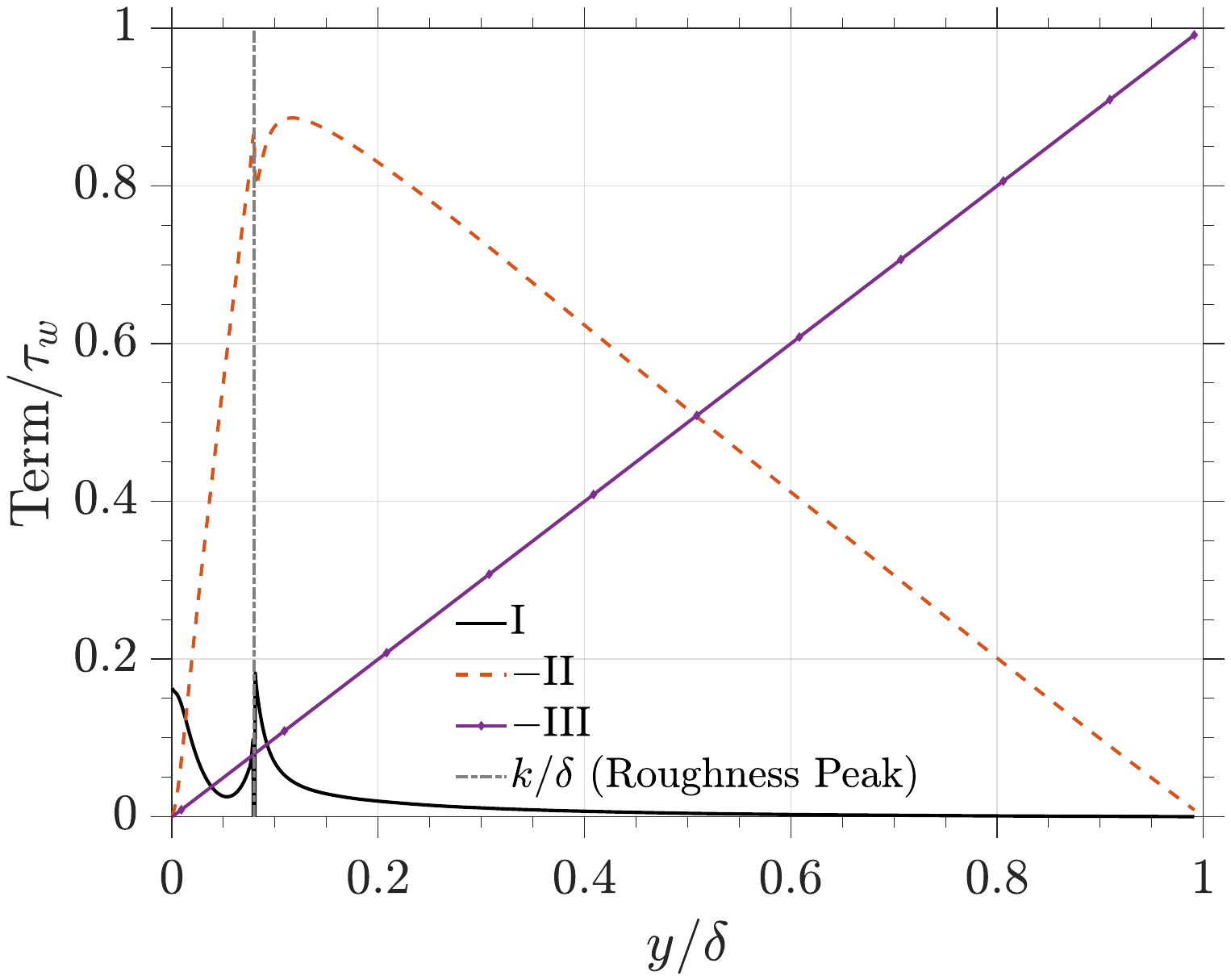}
    \caption{}
    \label{fig:comp-turb-channel-terms-rough}
\end{subfigure}
\begin{subfigure}{0.5\textwidth}
    \centering
    \captionsetup{justification=centering}
    \includegraphics[width=\textwidth]{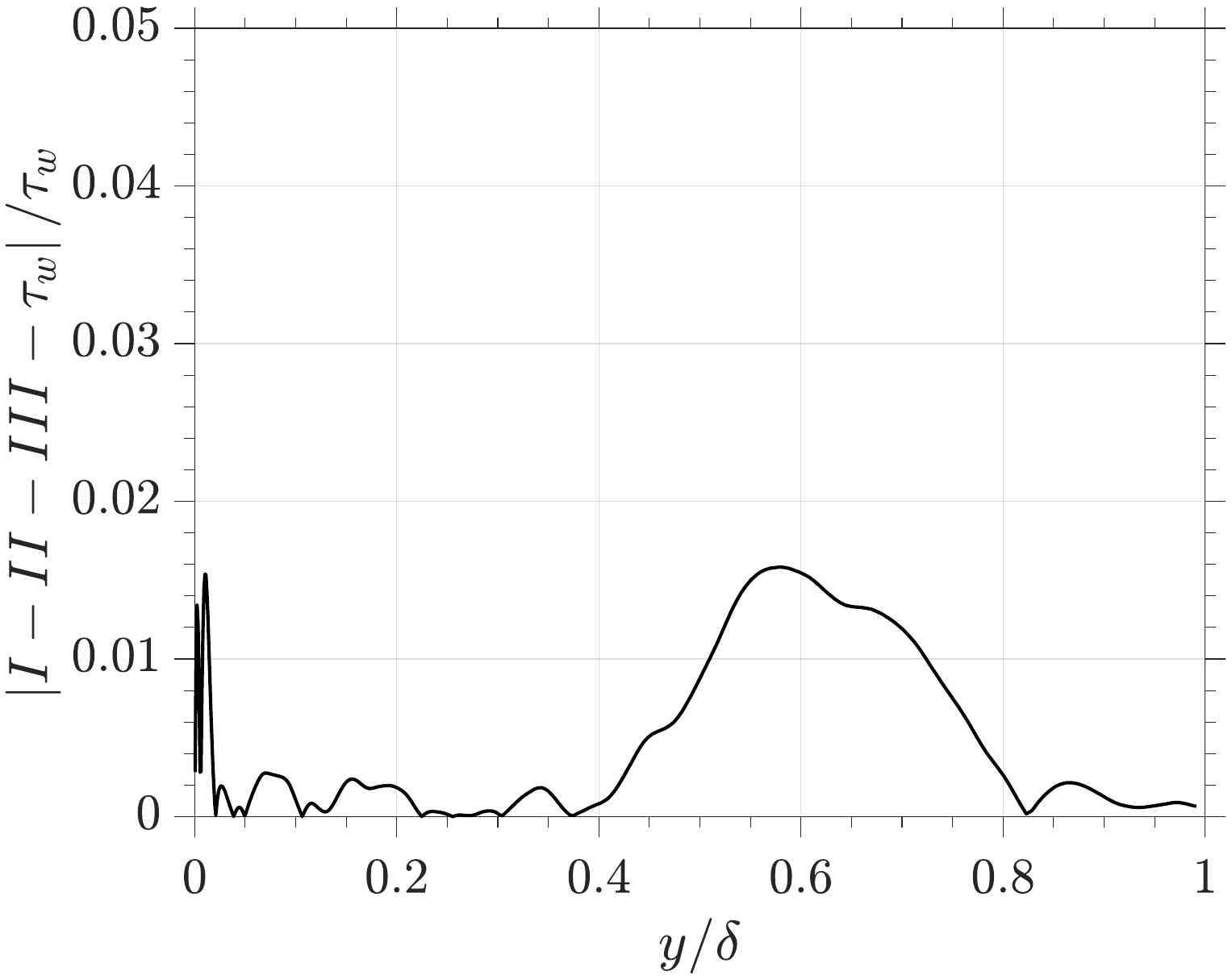}
    \caption{}
    \label{fig:comp-turb-channel-err-smooth}
\end{subfigure}%
~
\begin{subfigure}{0.5\textwidth}
    \centering
    \captionsetup{justification=centering}
    \includegraphics[width=\textwidth]{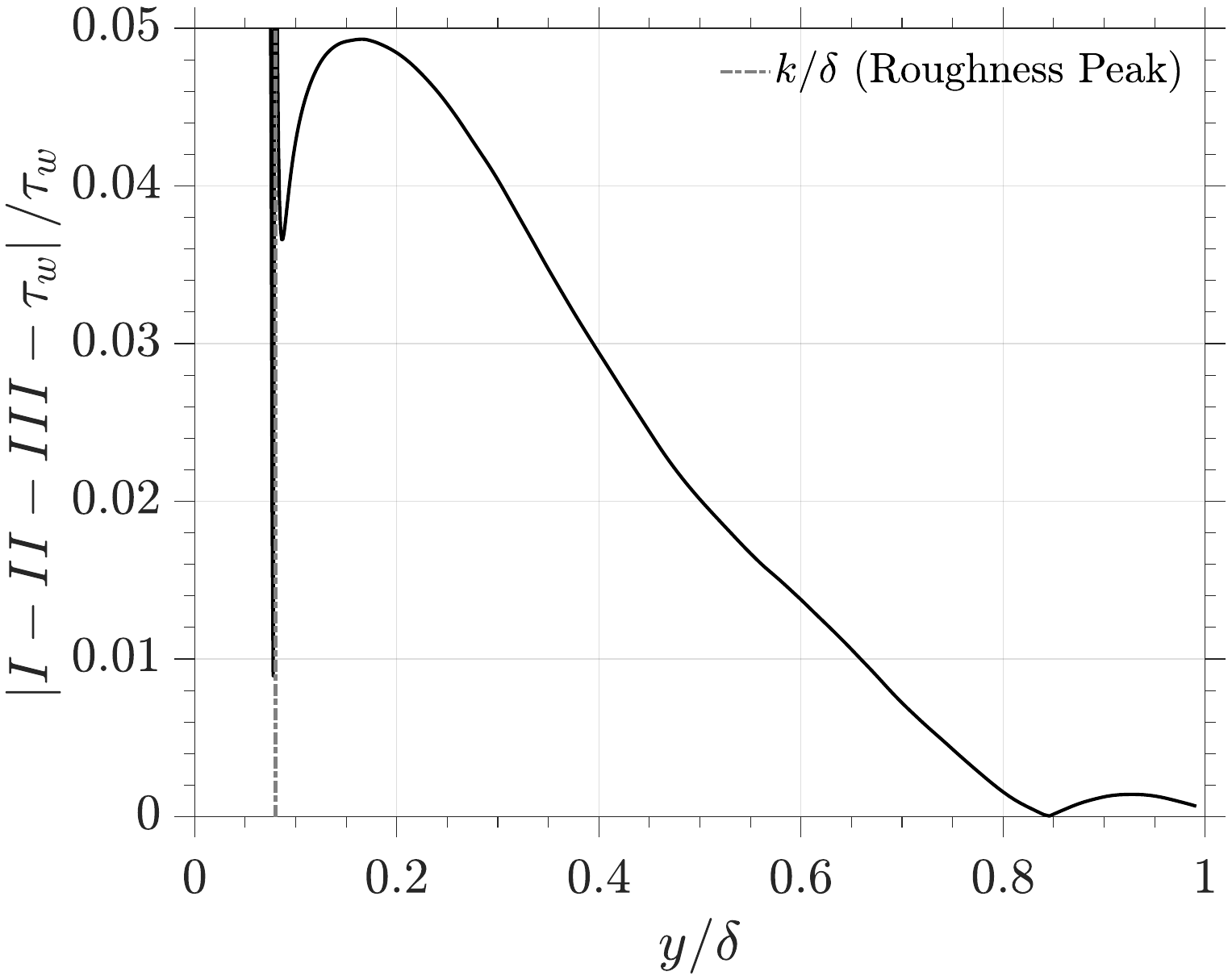}
    \caption{}
    \label{fig:comp-turb-channel-err-rough}
\end{subfigure}
\caption{Comparing smooth-wall to rough-wall turbulent channel flow at $M_b=4$ and $Re_\tau\approx1000$, wall-normal distance in outer scaling. Smooth wall (left), rough wall (right), contribution of stress term (top), error in wall shear stress estimate (bottom).}
\label{fig:comp-turb-channel-compare-Mb4-Ret1000}
\end{figure}

\begin{figure}[h!]
\centering
\captionsetup{justification=centering}
\begin{subfigure}{0.5\textwidth}
    \centering
    \captionsetup{justification=centering}
    \includegraphics[width=\textwidth]{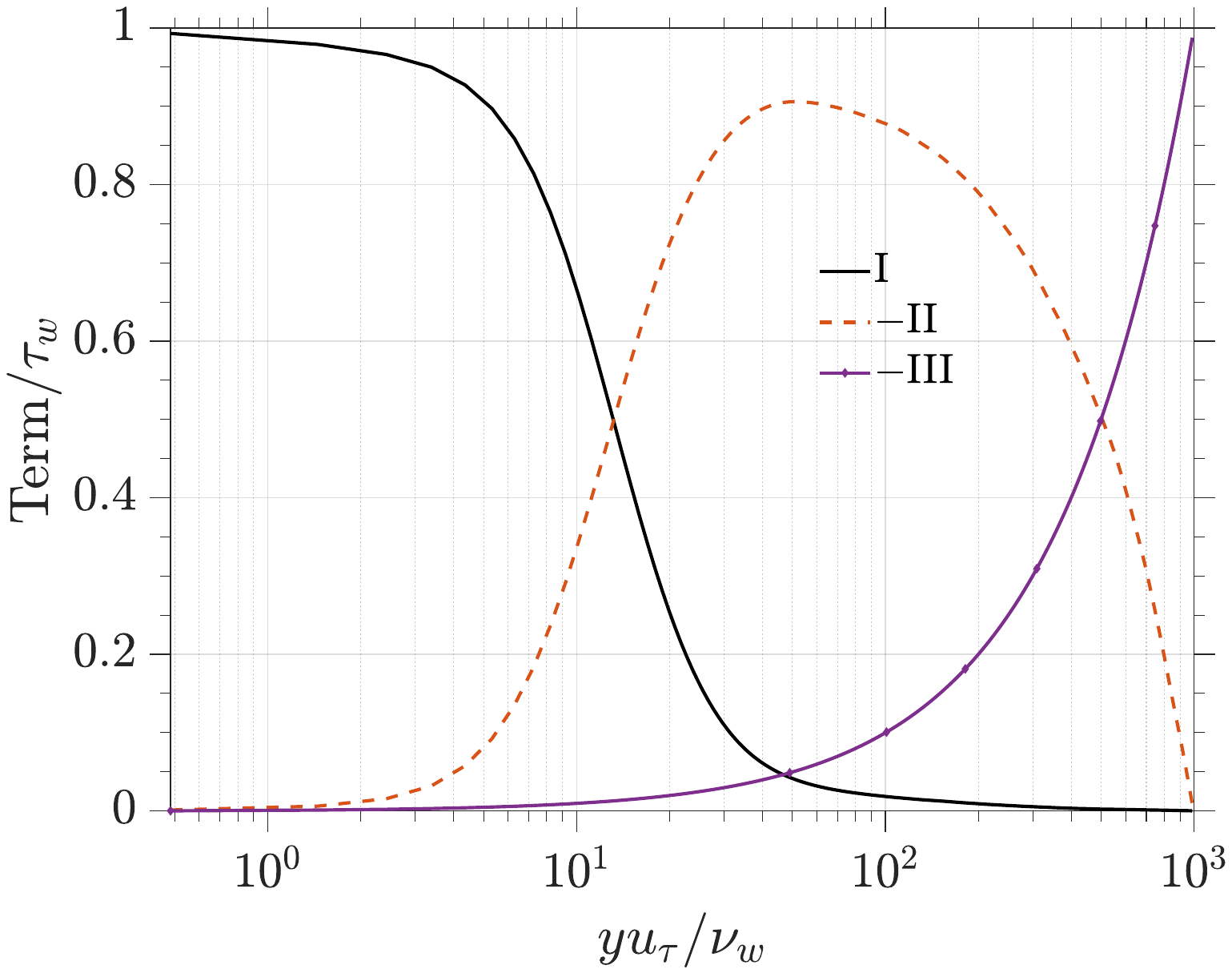}
    \caption{}
    \label{fig:comp-turb-channel-terms-plus-smooth}
\end{subfigure}%
~
\begin{subfigure}{0.5\textwidth}
    \centering
    \captionsetup{justification=centering}
    \includegraphics[width=\textwidth]{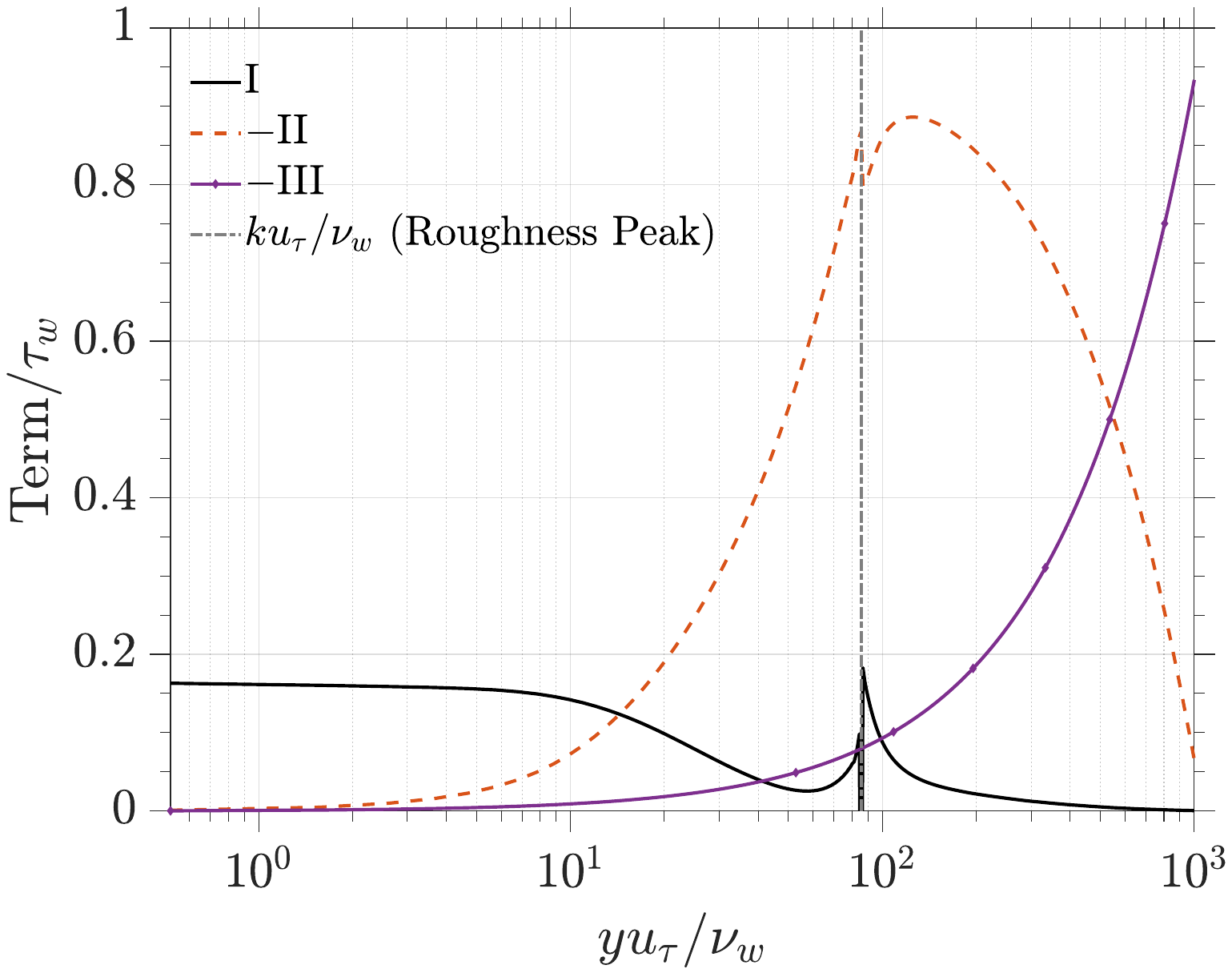}
    \caption{}
    \label{fig:comp-turb-channel-terms-plus-rough}
\end{subfigure}
\begin{subfigure}{0.5\textwidth}
    \centering
    \captionsetup{justification=centering}
    \includegraphics[width=\textwidth]{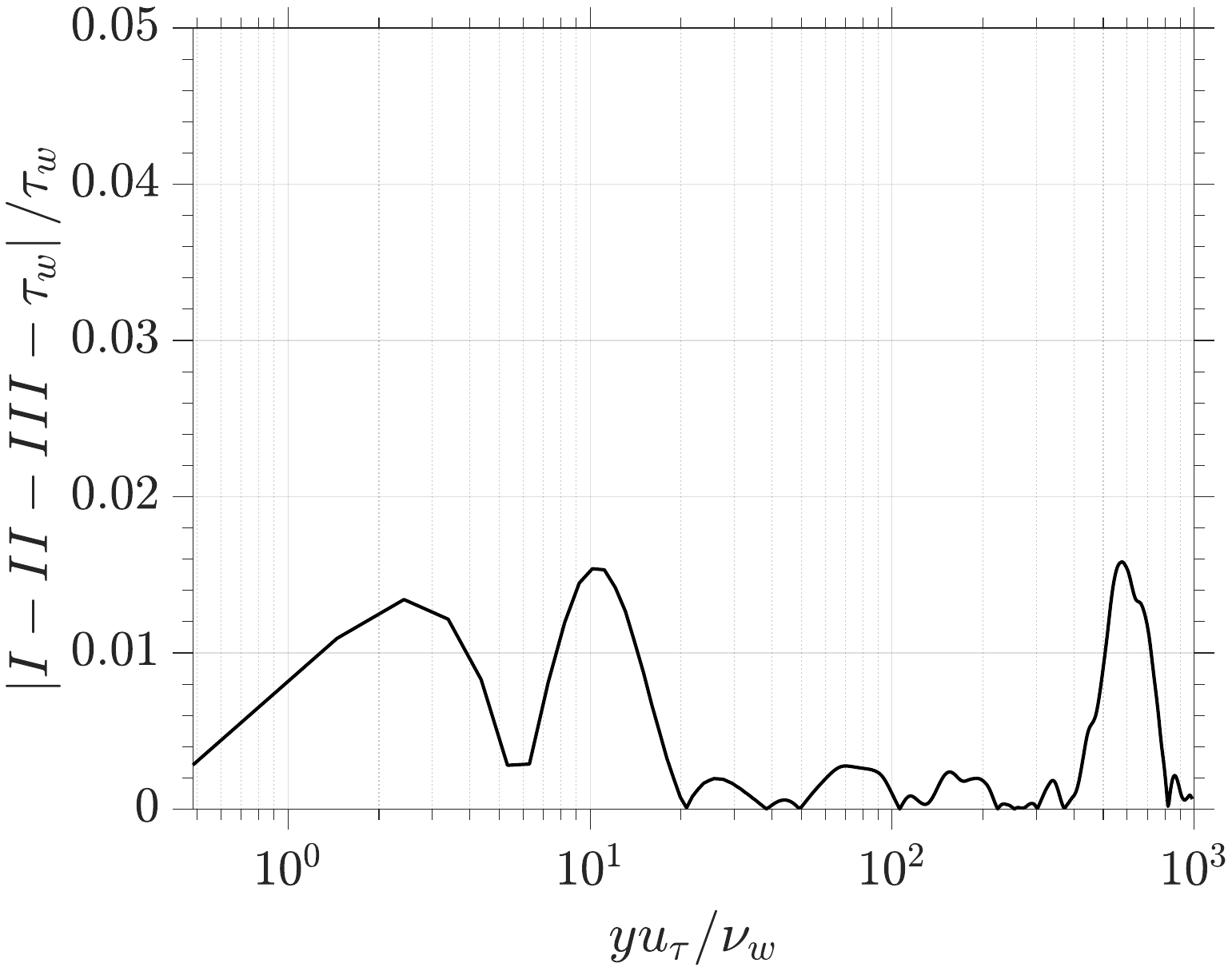}
    \caption{}
    \label{fig:comp-turb-channel-err-plus-smooth}
\end{subfigure}%
~
\begin{subfigure}{0.5\textwidth}
    \centering
    \captionsetup{justification=centering}
    \includegraphics[width=\textwidth]{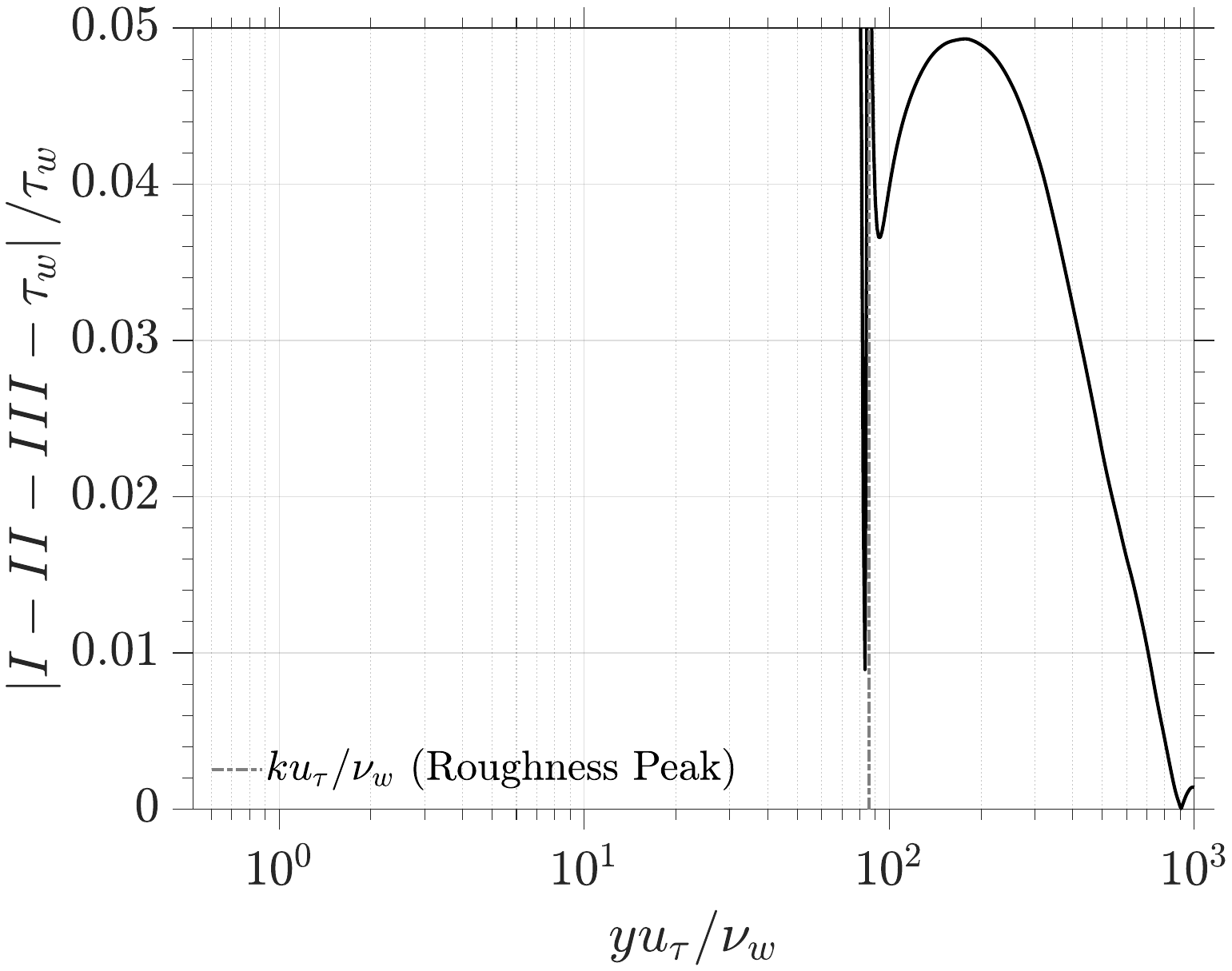}
    \caption{}
    \label{fig:comp-turb-channel-err-plus-rough}
\end{subfigure}
\caption{Comparing smooth-wall to rough-wall turbulent channel flow at $M_b=4$ and $Re_\tau\approx1000$, wall-normal distance in inner scaling. Smooth wall (left), rough wall (right), contribution of stress term (top), error in wall shear stress estimate (bottom).}
\label{fig:comp-turb-channel-compare-Mb4-Ret1000-plusunits}
\end{figure}

Extending from the previous demonstration, compressible rough-walled turbulent channel flow is the next test case. Modesti et al. \cite{Modesti22} perform DNS of supersonic turbulent channel flow over cubical roughness elements, spanning a range of bulk Mach numbers and roughness Reynolds numbers. The dataset comprises both smooth and rough wall cases, spanning friction Reynolds numbers 500$-$1000 and bulk Mach numbers $M_b = 0.3-4$ \cite{Modesti22_database}. The presence of the roughness alters the velocity profile in the viscous sublayer (near wall) and indicates that with transitional and fully rough roughness it is not possible to compute the wall shear stress only from the gradient of the average velocity profile near the wall. The present shear stress estimation method is capable of handling both transitional and fully rough regimes because the stress balance extends all the way to the channel centerline where an estimate can be taken from the cleaner data near the core of the channel. Figure \ref{fig:comp-turb-channel-compare-Mb4-Ret1000} shows the contribution of each stress term and the associated error in the estimated wall shear stress from the value provided in the dataset, for both the smooth and rough wall at the $M_b=4$ and $Re_\tau\approx1000$ conditions. Figure \ref{fig:comp-turb-channel-compare-Mb4-Ret1000-plusunits} shows the same comparison but with wall-normal distance reported with inner scaling. For the smooth wall, the maximum error in the estimated wall shear stress is less than 2\% and for the rough wall (beyond the roughness height) it is less than 5\%. Regardless of surface condition, the error in the estimate taken at the centerline is essentially zero.

Table \ref{tb:rough-channel-cases} tabulates all flow conditions provided in the dataset and includes a comparison of the wall shear stress between the present estimate and original value from the dataset. The value for the present estimate is taken as the value of the stress balance at the channel centerline. The data provided was nondimensional, so it was re-dimensionalized by setting the wall density $\rho_w = 0.1$ [kg/m$^3$] and wall temperature $T_w = 300$ [K]. Viscosity was evaluated through a power law with exponent 0.76, consistent with the original publication \cite{Modesti22}.

\begin{table}[h!]
\centering
\captionsetup{justification=centering}
\caption{Wall shear stress estimate comparison for all turbulent channel flow conditions provided in Modesti et al. \cite{Modesti22_database} dataset. Including smooth wall, transitional-roughness, and fully-rough regimes.}
\begin{tabular}{lcccccc}
\hline
Case & $M_b$ & $Re_\tau$ & $k^+$ & \makecell{$\tau_w$ [Pa]\\[-10pt] dataset} & \makecell{$\tau_w$ [Pa]\\[-10pt] present} & error [\%] \\\hline \hline
S03\_500 & 0.3 & 506 & N/A & 3.190 & 3.187 & -0.094 \\
S2\_500 & 2 & 488 & N/A & 75.287 & 75.202 & -0.113 \\
S4\_500 & 4 & 506 & N/A & 58.833 & 58.767 & -0.112 \\
S2\_1000 & 2 & 1003 & N/A & 61.806 & 61.774 & -0.052 \\
S4\_1000 & 4 & 1019 & N/A & 60.825 & 60.785 & -0.066 \\
R03\_500 & 0.3 & 513 & 41.0 & 10.061 & 10.049 & -0.113 \\
R2\_500 & 2 & 498 & 39.9 & 205.280 & 205.258 & -0.011 \\
R4\_500 & 4 & 518 & 41.5 & 163.573 & 163.568 & -0.003 \\
R2\_1000 & 2 & 1034 & 82.7 & 212.729 & 212.742 & 0.006 \\
R4\_1000 & 4 & 1072 & 85.8 & 185.907 & 185.783 & -0.067 \\ \hline
\end{tabular}
\label{tb:rough-channel-cases}
\end{table}

Modesti et al. \cite{Modesti22} use the following definitions: ``For roughness Reynolds numbers $k_s^+ \lesssim 5$ the flow is hydraulically smooth, that is the roughness does not induce any additional drag. As the roughness Reynolds number increases ($5 \lesssim k_s^+ \lesssim 80$) the flow becomes transitionally rough and in this regime both viscous and pressure drag are important. For $k_s^+ \gtrsim 80$ the flow becomes fully rough, meaning the skin-friction coefficient does not depend on the Reynolds number.'' With Table \ref{tb:rough-channel-cases}, one can conclude that the present method works well regardless if the flow is in the hydrodynamically smooth, transitional, or fully rough regime.

\subsection{Mach 2.9, ZPG, Rough-Wall, Turbulent Flat Plate}\label{sec:Mach 2.9, ZPG, Rough-Wall, Turbulent Flat Plate}
To further explore rough-wall conditions, a zero-pressure-gradient flat plate, rough-walled, turbulent boundary layer has been included in the demonstration cases. We performed a DNS of a turbulent boundary layer over sinusoidal surface roughness using US3D, the unstructured finite volume Navier-Stokes solver developed at the University of Minnesota \cite{candler15_US3D}. The flow conditions, computational grid, and roughness profile are chosen to follow the DNS of Muppidi and Mahesh \cite{Muppidi12}, originally studying laminar to turbulent transition due to a patch of distributed surface roughness for a Mach 2.9 supersonic flat plate boundary layer flow. A difference between the present simulation and the original work by Muppidi and Mahesh is that we continue the roughness profile all the way to the outflow of the domain. 

\begin{figure}[h!]
    \centering
    \captionsetup{justification=centering}
    \includegraphics[width=0.4\linewidth]{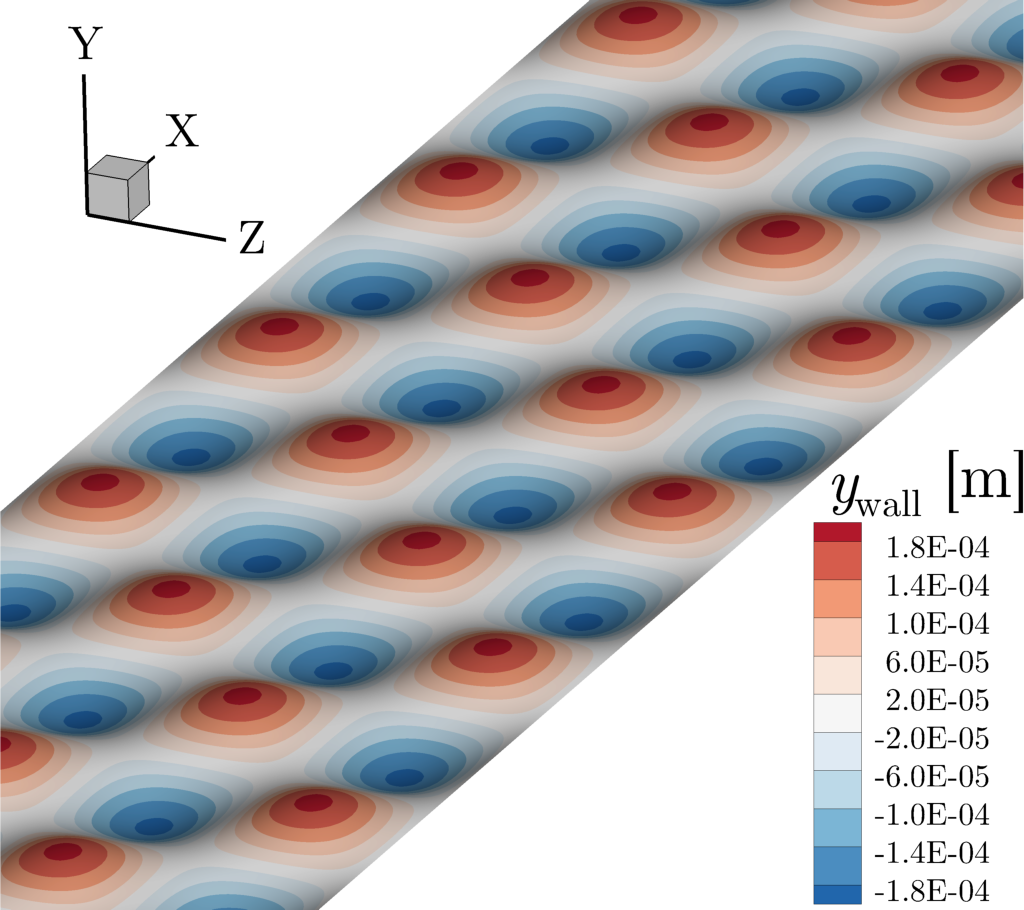}
    \caption{Sinusoidal surface profile for Mach 2.9, ZPG, rough-wall, turbulent flat plate.}
    \label{fig:wavy-surface-zpg-rough-bl}
\end{figure}

\begin{figure}[h!]
    \centering
    \captionsetup{justification=centering}
    \includegraphics[width=0.9\linewidth]{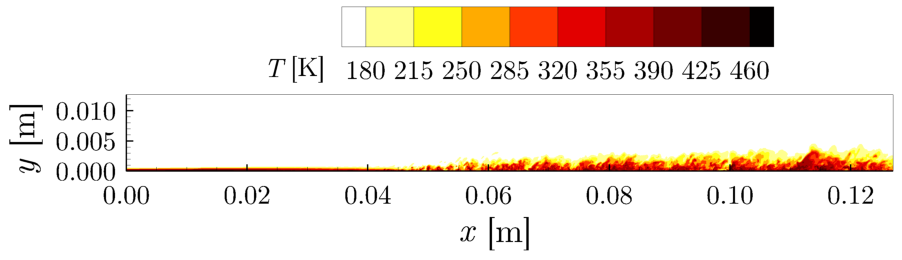}
    \caption{Instantaneous static temperature contours of Mach 2.9, ZPG, rough-wall, turbulent flat plate.}
    \label{fig:x-y-slice-t-zpg-rough-bl}
\end{figure}

The specific details of the present simulations are as follows. The length ($x$, streamwise), width ($z$, spanwise), and height ($y$, wall-normal) of the domain are $L_x=0.127$ [m], $L_z=0.004445$ [m], and $L_y=0.0127$ [m] respectively. The corresponding number of cells are $N_x=2000$, $N_z=94$, and $N_y=192$. The first cell off the wall has a wall-normal spacing of 	$\Delta y_w=1.27\times{10}^{-6}$ [m], with tanh stretching away from the wall. The surface roughness profile is prescribed according to Eq. \ref{eq:Muppidi-roughness-profile}, where $k=0.0001905$ [m] is the roughness amplitude, $\kappa_x=100\pi/L_x$ is the streamwise wavenumber, and $\kappa_z=4\pi/L_z$ is the spanwise wavenumber. Roughness begins 0.0127 [m] from a prescribed inflow plane and then persists to the end of the domain at a supersonic outflow. The freestream conditions are $M_\infty=2.9$, $T_\infty= 170$ [K], $\rho_\infty=3.8014\times10^{-1}$ [kg/m$^3$], $U_\infty=757.927$ [m/s], and $Re_\infty=2.3\times{10}^7$ [1/m]. A compressible Blasius boundary layer is prescribed at the inflow ($x$-location for boundary layer similarity solution is $x_{BL}=0.1016$ [m]). The wall boundary conditions are no-slip, no-penetration, and isothermal with $T_w=455.94$ [K]. The spanwise boundaries of the domain are periodic. The fluid is perfect-gas air, and Sutherland's law was used for viscosity. A fourth-order kinetic energy consistent (KEC) flux scheme from Subbareddy and Candler \cite{subbareddy09} and second-order implicit time integration was used for the DNS. A three-dimensional view of the surface roughness profile is shown in Fig. \ref{fig:wavy-surface-zpg-rough-bl}, and an instantaneous snapshot of temperature for an $x-y$ slice of the domain is shown in Fig. \ref{fig:x-y-slice-t-zpg-rough-bl} to illustrate the domain geometry and turbulence.

\begin{equation}
y_{\text{wall}}=k\sin\left(\kappa_x x\right)\sin\left(\kappa_z z\right)
\label{eq:Muppidi-roughness-profile}
\end{equation}

\begin{figure}[h!]
\centering
\captionsetup{justification=centering}
\begin{subfigure}{0.5\textwidth}
    \centering
    \captionsetup{justification=centering}
    \includegraphics[width=\textwidth]{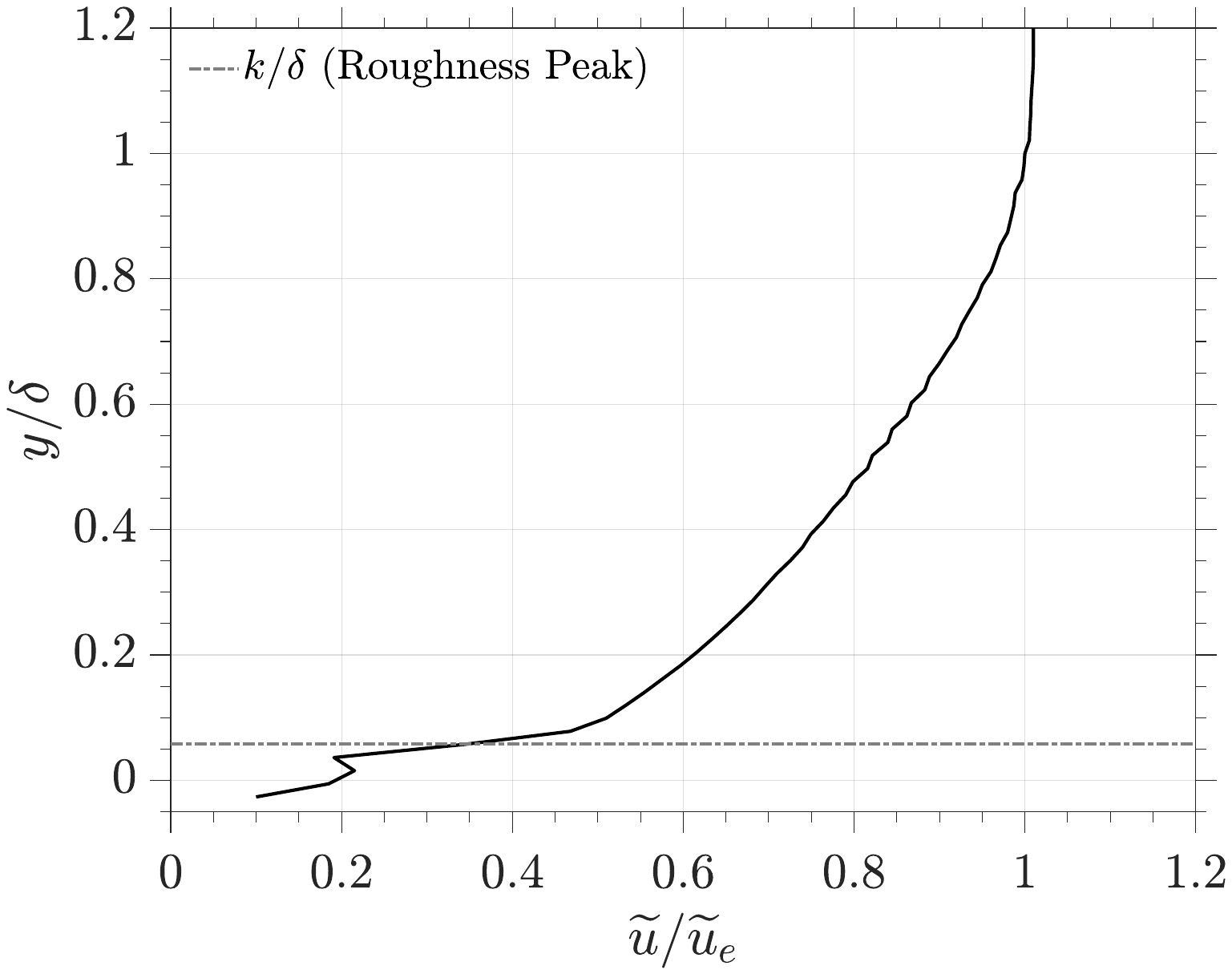}
    \caption{Outer scaling}
    \label{fig:u_ue-zpg-rough-bl}
\end{subfigure}%
~
\begin{subfigure}{0.5\textwidth}
    \centering
    \captionsetup{justification=centering}
    \includegraphics[width=\textwidth]{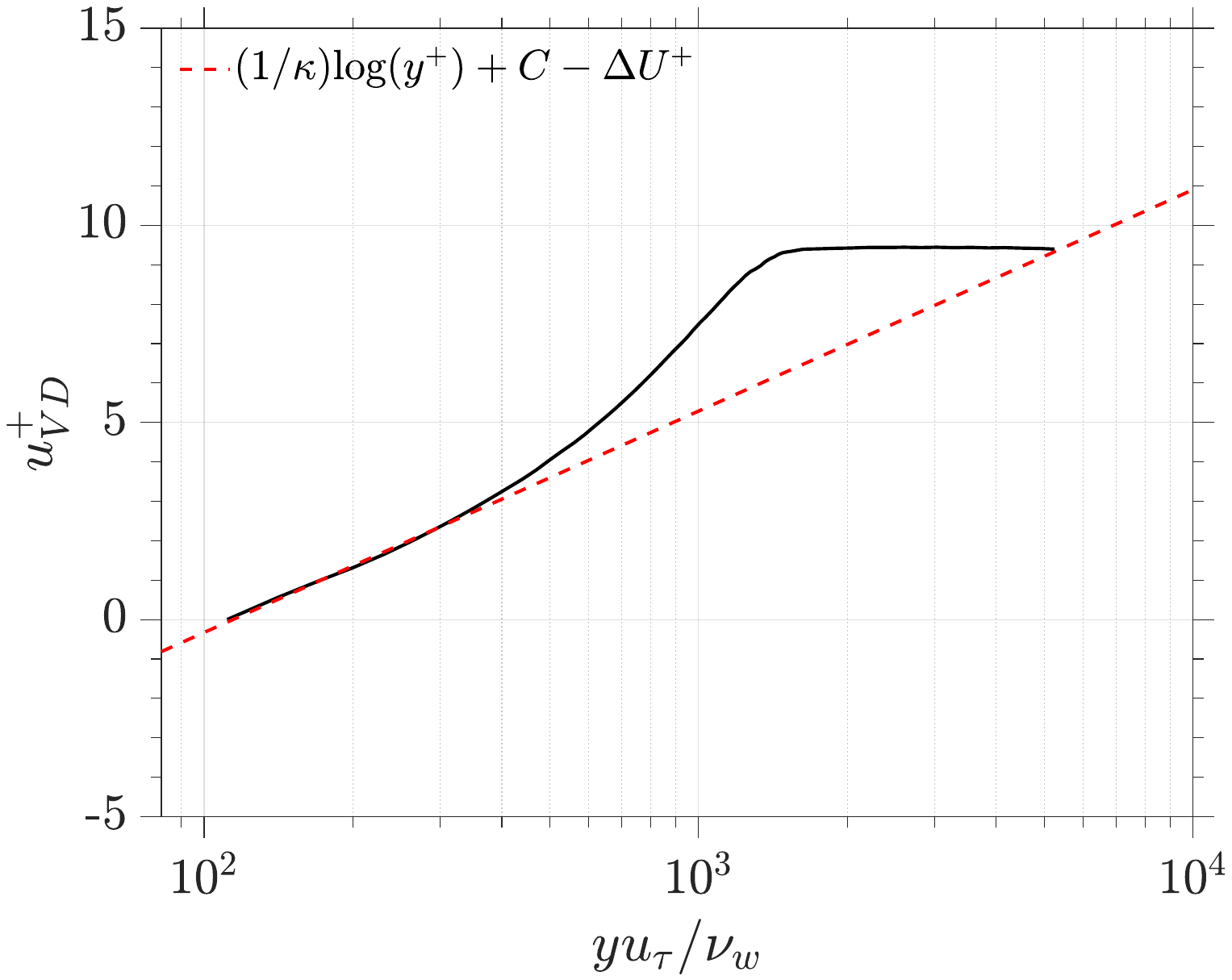}
    \caption{Van Driest inner scaling}
    \label{fig:uplus-zpg-rough-bl}
\end{subfigure}
\caption{Time and span averaged mean flow profile for Mach 2.9, ZPG, rough-wall, turbulent flat plate. For (b), $\kappa=0.41$, $C=5.2$, $\Delta U^+=16.76$, and data points below the roughness peak are truncated.}
\label{fig:mean-flow-profiles-zpg-rough-bl}
\end{figure}

The flow is allowed to transition due to the presence of the roughness. In the turbulent downstream region, a spanwise-wall-normal slice at $x_q=0.1$ [m] from the inflow was time averaged for 24 flow through times based on edge conditions, and then span averaged to obtain the wall-normal profiles needed in the shear stress integral relation. A ``true'' value of the shear stress was obtained from the Clauser fit technique \cite{Clauser54}. Fitting in the range $100\leq y^+\leq350$ resulted in $u_\tau=49.616$ [m/s], using $\rho_w$ at the roughness peak subsequently led to $\tau_w=437.35$ [Pa]. The mean flow profile is illustrated in Fig. \ref{fig:mean-flow-profiles-zpg-rough-bl}.

\begin{figure}[h!]
\centering
\captionsetup{justification=centering}
\begin{subfigure}{0.5\textwidth}
    \centering
    \captionsetup{justification=centering}
    \includegraphics[width=\textwidth]{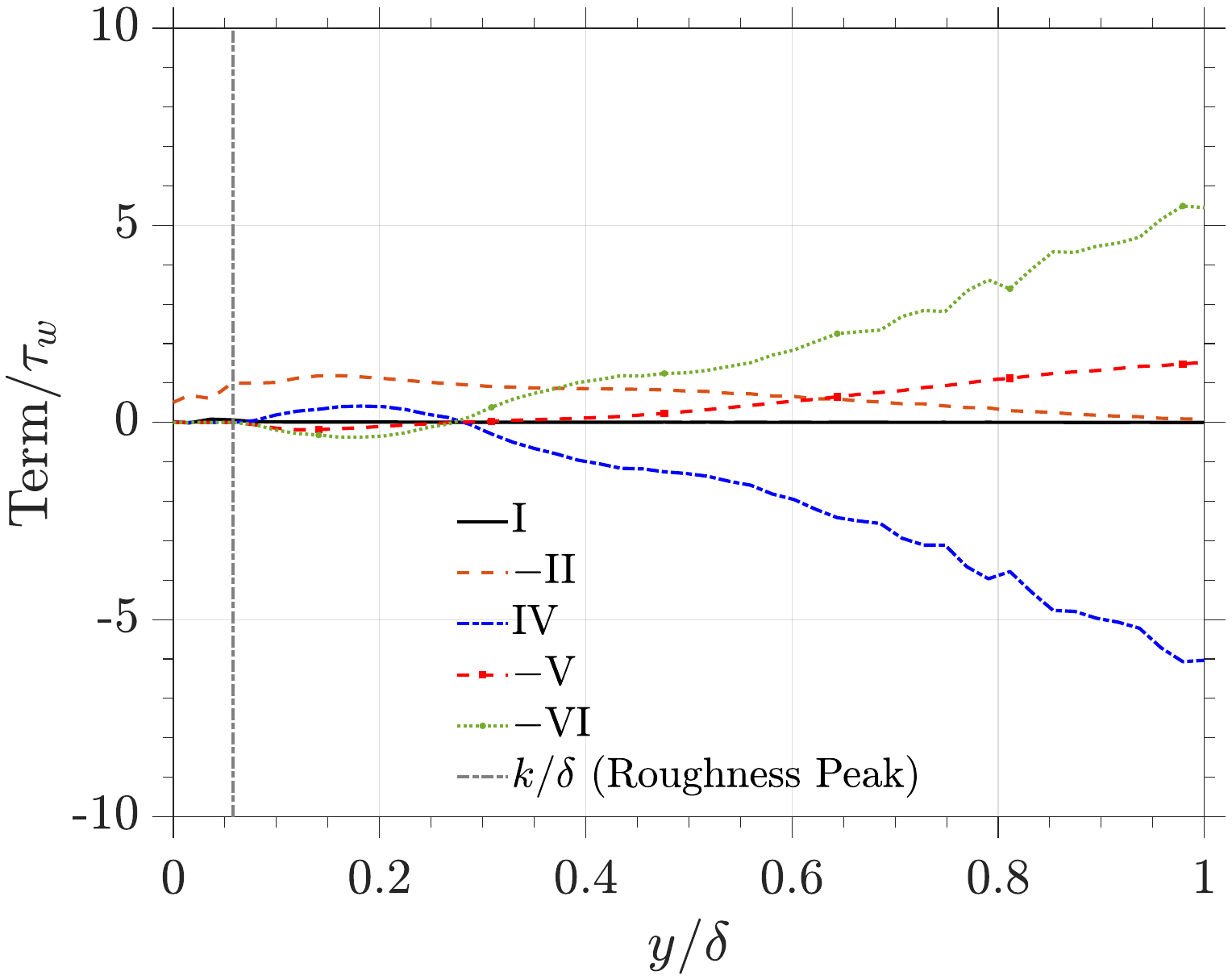}
    \caption{Contribution of stress term}
    \label{fig:zpg-rough-bl-bb-terms}
\end{subfigure}%
~
\begin{subfigure}{0.5\textwidth}
    \centering
    \captionsetup{justification=centering}
    \includegraphics[width=\textwidth]{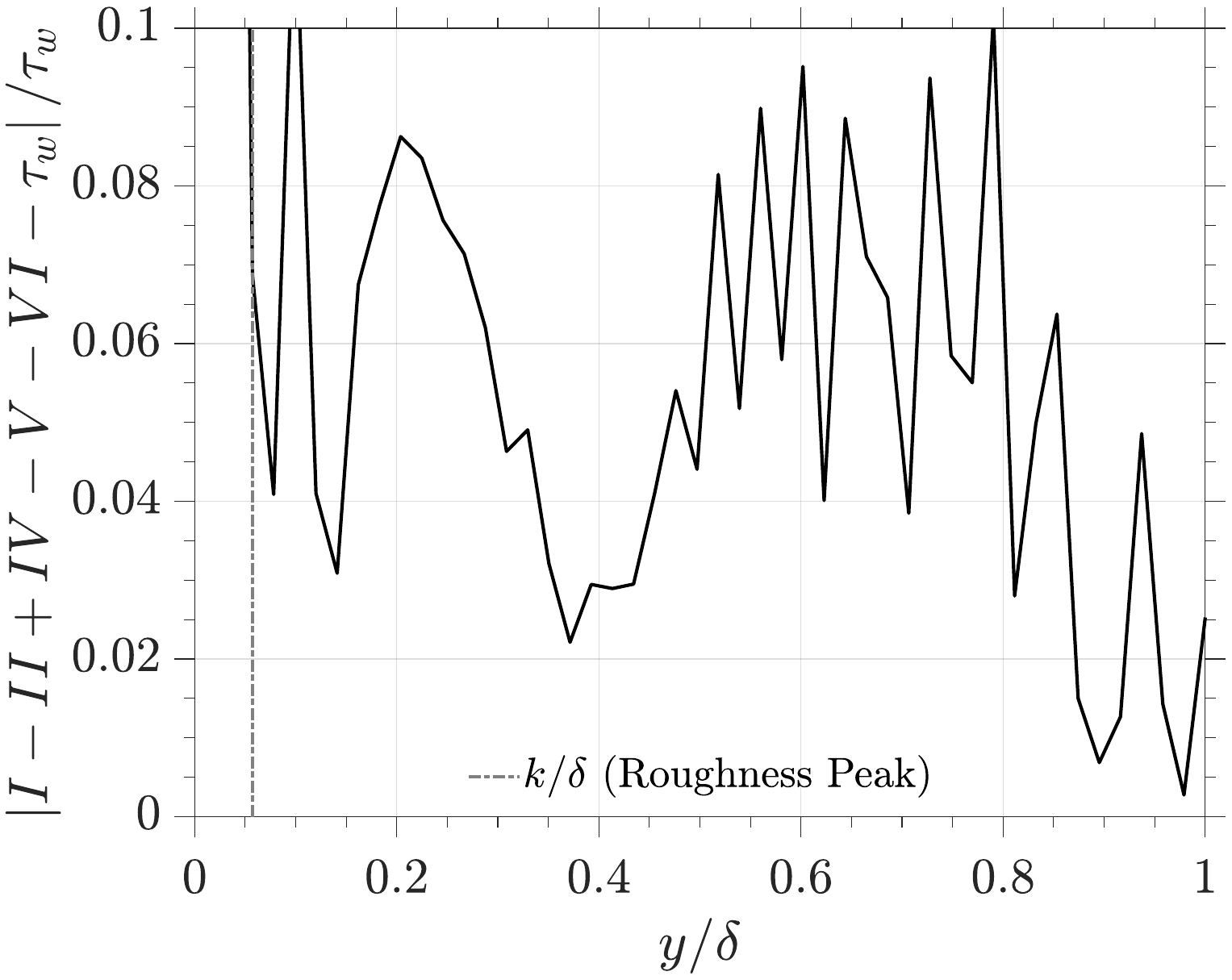}
    \caption{Error}
    \label{fig:zpg-rough-bl-bb-err}
\end{subfigure}
\caption{Mach 2.9, ZPG, rough-wall, turbulent flat plate contribution of each stress term and error compared to Clauser fit shear stress as a function of wall-normal distance.}
\label{fig:zpg-rough-bl-terms}
\end{figure}

Figure \ref{fig:zpg-rough-bl-terms} shows the contribution of each stress term and the associated error in the estimated wall shear stress as compared to the wall shear stress computed form the Clauser fit of the mean velocity profile. The average wall shear stress from momentum equation balance (above the roughness peak) is $\tau_w=437.46$ [Pa], recalling the wall shear stress from the Clauser fit, $\tau_{w,Clauser}=437.35$ [Pa], the average percent error in wall shear stress estimate is therefore 0.03\%. At no point in the boundary layer does the local error go above 10\%. The local error is larger than some of the other cases shown, but there are a few considerations that still makes this a compelling result. First, this case relies on streamwise gradients computed from actual DNS data (not from a similarity solution) which is affected by the averaging window. Second, with the presence of the roughness, selecting a virtual origin is necessary in order to integrate terms IV, V, and VI.

\begin{figure}[h!]
\centering
\captionsetup{justification=centering}
\begin{subfigure}{0.5\textwidth}
    \centering
    \captionsetup{justification=centering}
    \includegraphics[width=\textwidth]{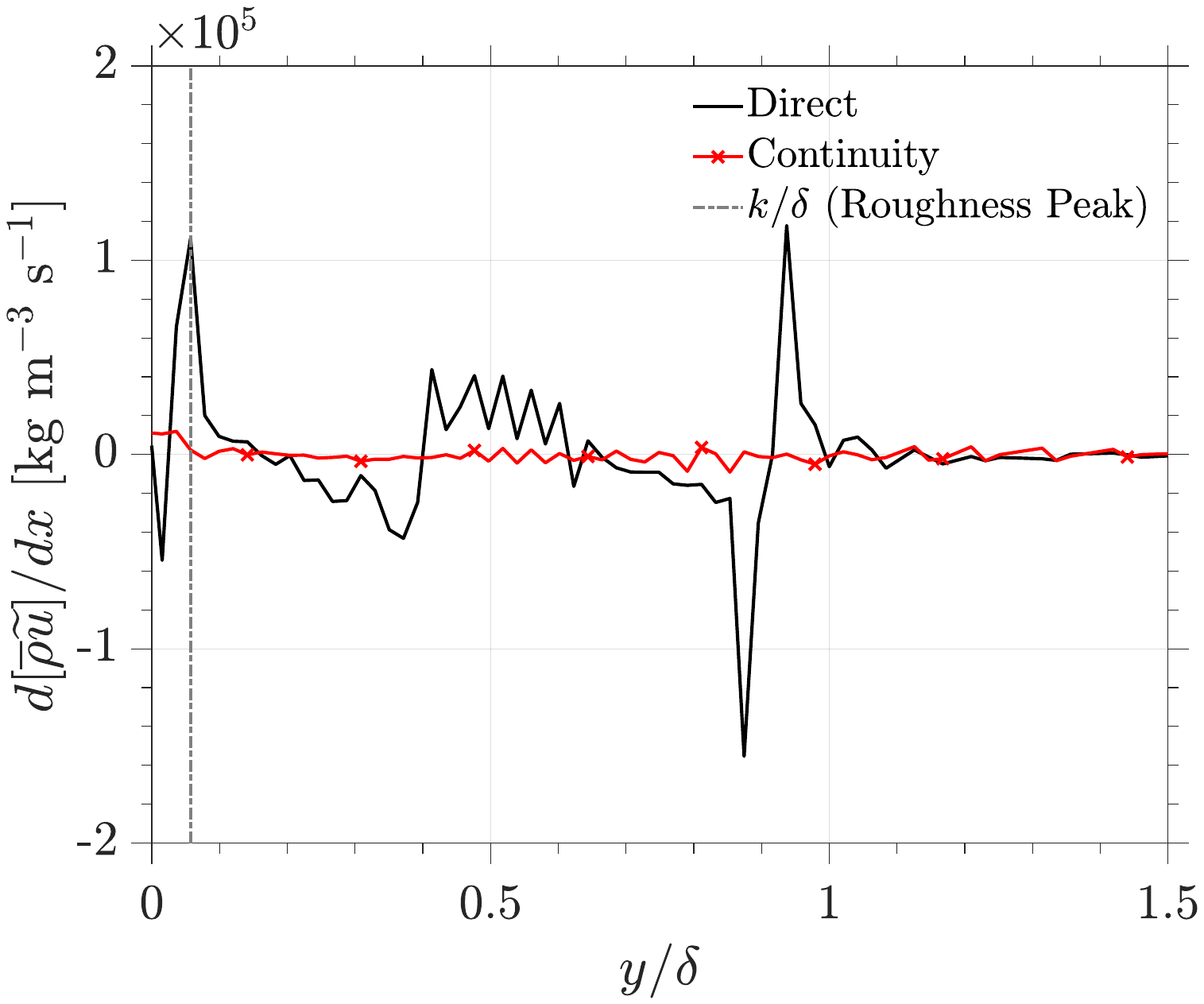}
    \caption{Streamwise mass flux gradient}
    \label{fig:zpg-rough-bl-dru_dx}
\end{subfigure}%
~
\begin{subfigure}{0.5\textwidth}
    \centering
    \captionsetup{justification=centering}
    \includegraphics[width=\textwidth]{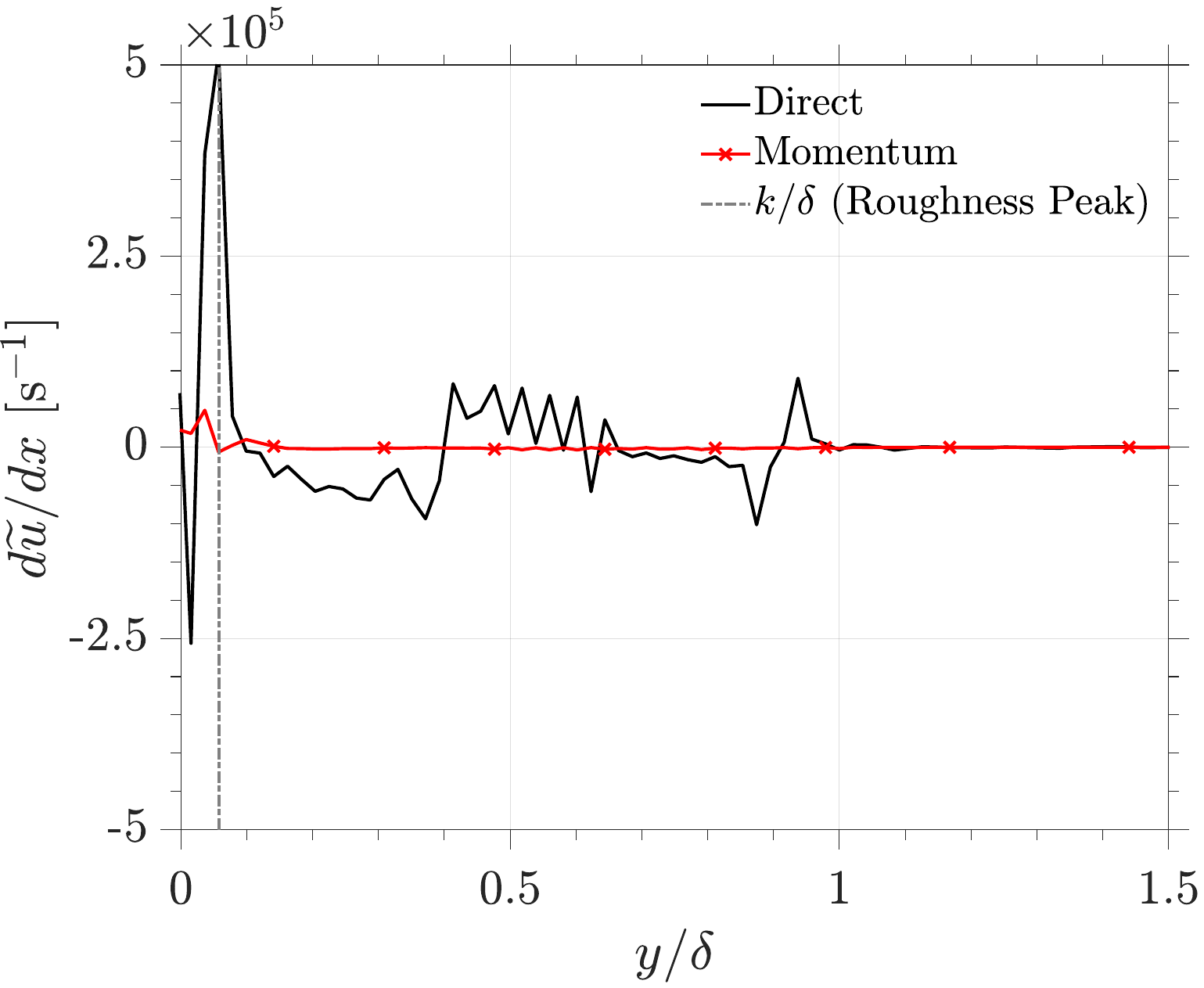}
    \caption{Streamwise velocity gradient}
    \label{fig:zpg-rough-bl-du_dx}
\end{subfigure}
\caption{Mach 2.9, ZPG, rough-wall, turbulent flat plate streamwise gradients comparison.}
\label{fig:zpg-rough-bl-str-gradients}
\end{figure}

Addressing the larger local error and sensitivity to the streamwise gradients, Fig. \ref{fig:zpg-rough-bl-str-gradients} compares the streamwise gradient values for both the mass flux and velocity when computed directly versus computed with Eqs. \ref{eq:continuity-gradient-swap} and \ref{eq:momentum-gradient-swap}. The ``Direct'' gradients are computed with a finite difference between two wall-normal profiles spaced in the streamwise direction by $6.35\times10^{-5}$ [m] (one cell downstream in the DNS grid). The ``Continuity'' and ``Momentum'', wall-normal gradient only, form comes from Eq. \ref{eq:continuity-gradient-swap} and Eq. \ref{eq:momentum-gradient-swap}, respectively. The direct calculation of the gradients are very sensitive to the grid resolution, time averaging, and span averaging. It is clear from this comparison that even for a full developed, ZPG boundary layer the direct calculation leads to spurious gradient values that are orders of magnitude larger than the expected result -- highlighting the benefit of the continuity and momentum equation gradient substitutions. However, even with the continuity and momentum substituted gradient forms, there is still some variation in the calculated streamwise gradient as a function of the wall normal direction and this explains the larger local error seen in Fig. \ref{fig:zpg-rough-bl-bb-err}.

At the given conditions and location, the roughness Reynolds number is $k^+=ku_\tau/\nu_w=81.94$ indicating a fully rough regime. The presence of large roughness significantly alters the behavior of the flow in the near wall region. Relevant to the present analysis, terms IV, V, and VI are significantly affected by the selection of a virtual origin for the zero location of the wall-normal integrations. For this case, the virtual origin was selected at the top of the roughness peak $y=k$. Including data below the roughness peak in the calculation of the integral terms led to average errors in the wall shear stress over 200\% from the Clauser fit. Moreover, the summation of the terms from an equivalent plot to Fig. \ref{fig:zpg-rough-bl-bb-terms} was not constant across the boundary layer height.

\subsection{Mach 6.0, Laminar, Hypersonic Blunt Body}
\begin{figure}[h!]
    \centering
    \captionsetup{justification=centering}
    \begin{subfigure}{0.5\textwidth}
        \centering
        \captionsetup{justification=centering}
        \includegraphics[width=\linewidth]{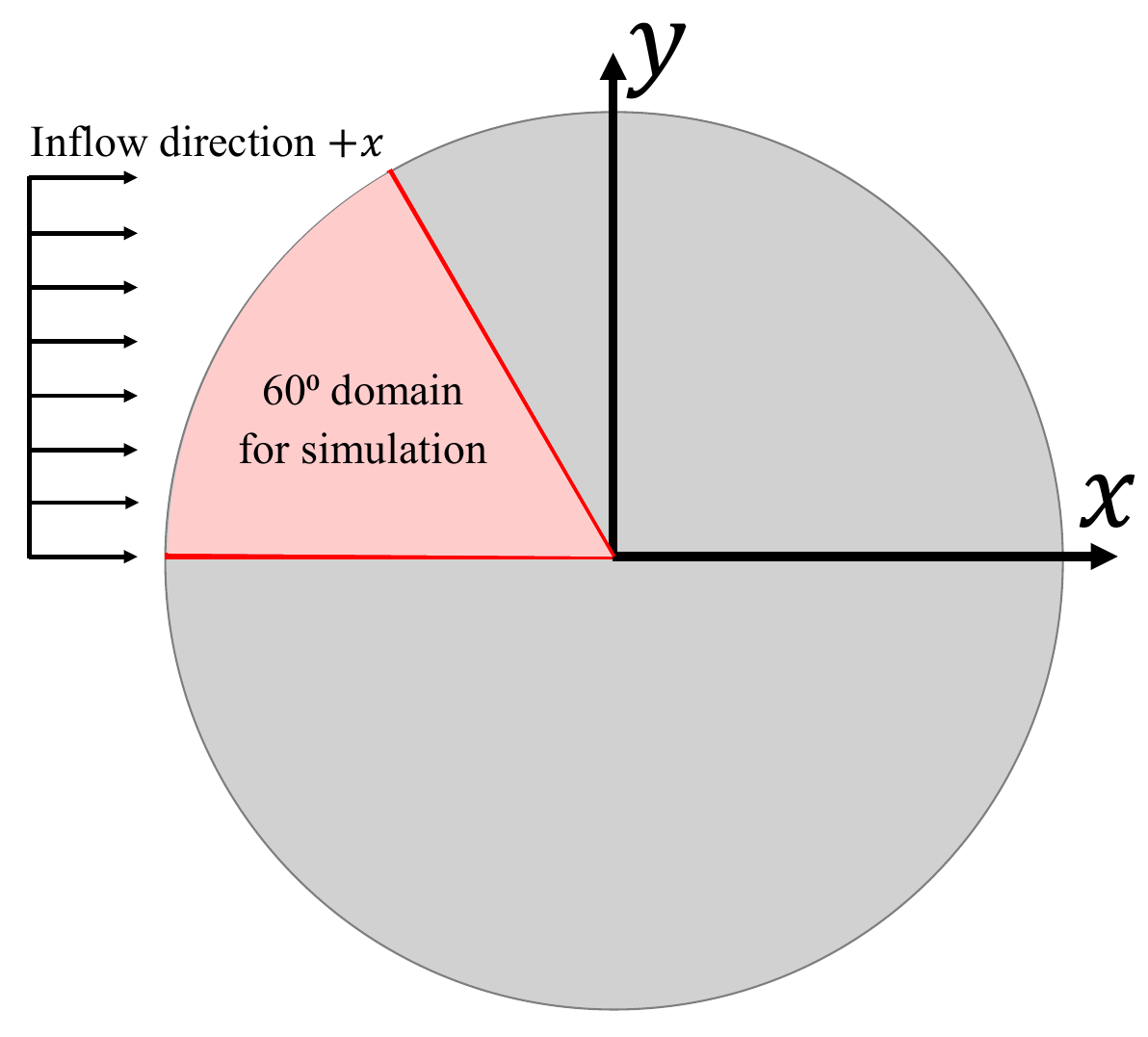}
        \caption{Cylinder geometry}
        \label{fig:cylinder-geometry}         
    \end{subfigure}%
    ~    
    \begin{subfigure}{0.5\textwidth}
        \centering
        \captionsetup{justification=centering}
        \includegraphics[width=\linewidth]{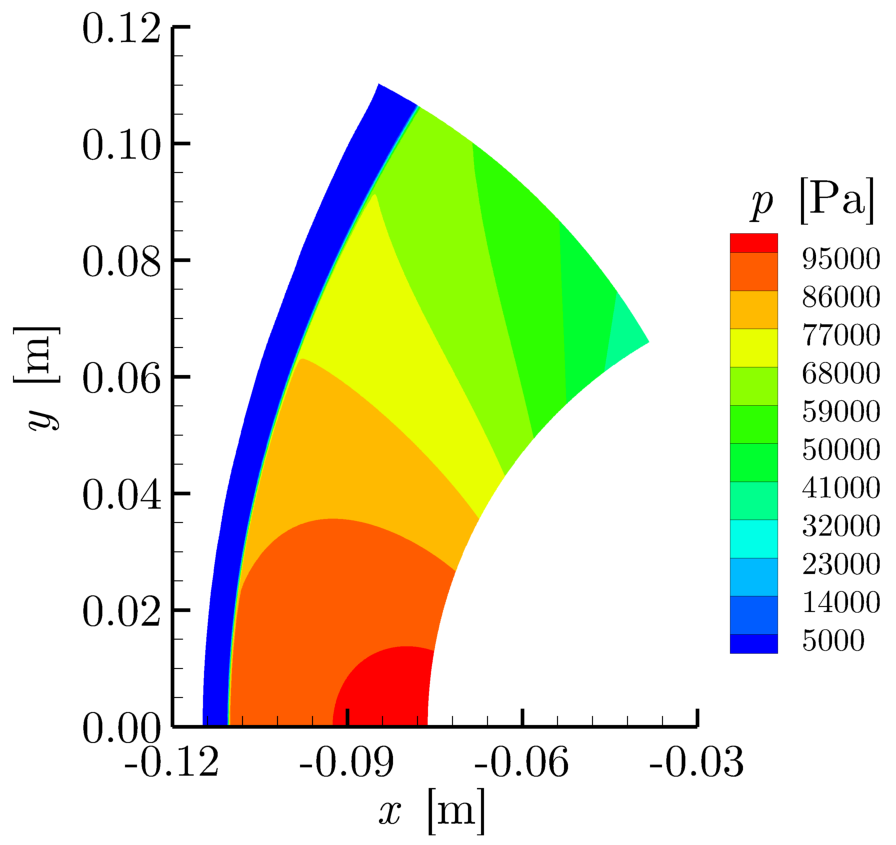}
        \caption{Static pressure contours}
        \label{fig:cylinder-P-contours}         
    \end{subfigure}
\caption{Two-dimensional cylinder simulation setup and solution. Origin of cylinder located at $x=0$ [m], $y=0$ [m].}
\label{fig:hypersonic-blunt-body-setup}
\end{figure}

Another demonstration that we simulated is a hypersonic blunt body, specifically a 0.1524 [m] smooth-walled cylinder in Mach 6 flow. This simulation was also performed using US3D. The boundary conditions are selected based on the conditions of Test 6975 from Hollis \cite{hollis17_NASATM}. The cylinder wall is an isothermal no-slip wall at $T_w = 300$ [K], and the freestream boundary conditions are as follows: $Re_\infty=2.74\times10^7$ [1/m], $T_\infty=58.6$ [K], $\rho_\infty=1.249\times 10 ^{-1}$ [kg/m$^3$], and $U_\infty=918.1$ [m/s]. The laminar solution of the compressible Navier-Stokes equations are solved numerically with implicit Euler time integration, second order inviscid fluxes in space, and modified Steger-Warming flux-vector splitting. The computational mesh is two-dimensional (one cell wide in the spanwise direction) and has 554 streamwise and 600 wall normal cells. Making use of symmetry, the domain extends from the stagnation line to a supersonic outflow 60\textdegree{} from the stagnation line. In this case because of the curved geometry, the wall shear stress is estimated based on the locally wall-parallel velocity component $U_{||}$ and the wall normal distance $\eta$. The maximum wall spacing in the streamwise direction, $s$ is $\Delta s_w^+<70$, and the maximum spacing in the wall normal direction is $\Delta \eta_w^+<0.4$. Fig. \ref{fig:hypersonic-blunt-body-setup} illustrates the cylinder geometry.

\begin{figure}[h!]
\centering
\captionsetup{justification=centering}
\begin{subfigure}{0.5\textwidth}
    \centering
    \captionsetup{justification=centering}
    \includegraphics[width=\textwidth]{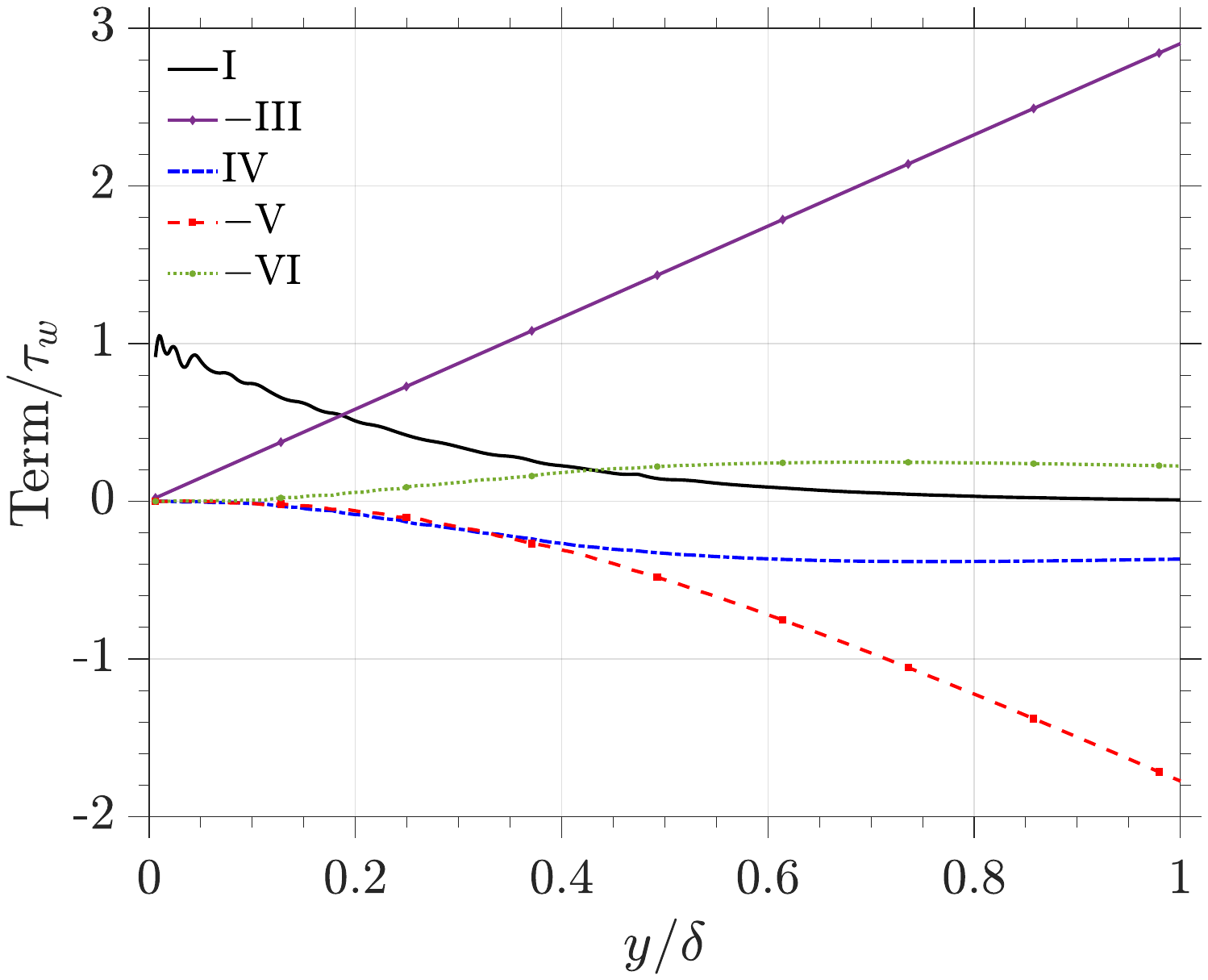}
    \caption{Contribution of stress term}
    \label{fig:hypersonic-bb-terms}
\end{subfigure}%
~
\begin{subfigure}{0.5\textwidth}
    \centering
    \captionsetup{justification=centering}
    \includegraphics[width=\textwidth]{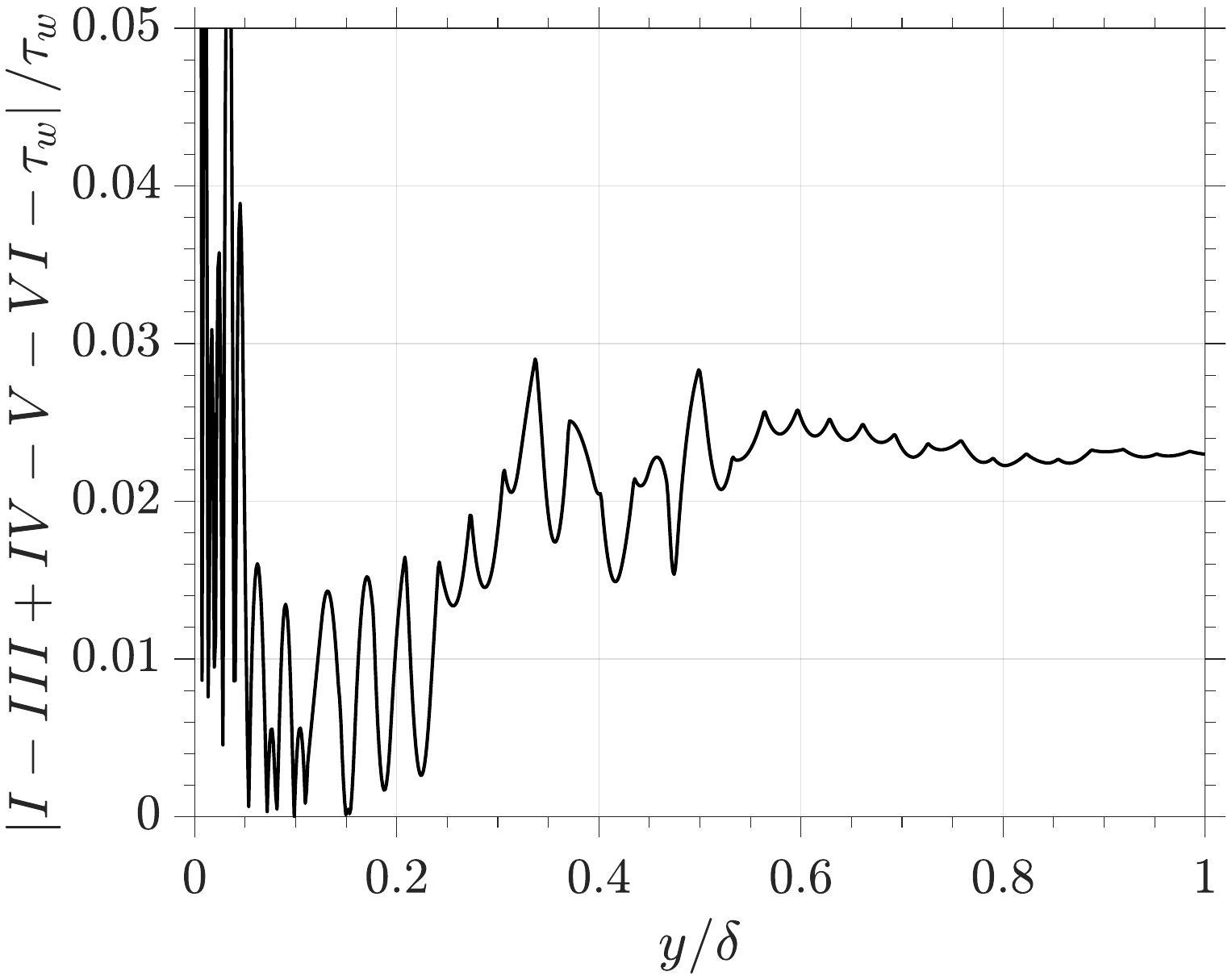}
    \caption{Error}
    \label{fig:hypersonic-bb-err}
\end{subfigure}
\caption{Hypersonic blunt body contribution of each stress term and error compared to wall shear stress computed from velocity gradient at the wall as a function of wall-normal distance.}
\label{fig:hypersonic-blunt-body-terms}
\end{figure}

For the demonstration case, wall normal lines are taken at 45\textdegree{} from the stagnation line. Figure \ref{fig:hypersonic-blunt-body-terms} shows the contribution of each stress term and the associated error in the estimated wall shear stress as compared to the wall shear stress computed from $\tau_w=\mu\frac{dU_{||}}{d\eta}$. Initially, there is large error associated to the computed gradients near the wall; however, the initial error does not contaminate the wall-normal integrals and by $y/\delta>0.05$ the difference does not exceed 3\% for the remainder of the boundary layer. In fact, if an average of the predicted wall shear stress from the integral relationship at all wall-normal locations in the boundary layer is taken, it results in $\tau_w=103.34$ [Pa], compared to $\tau_w=\mu\frac{dU_{||}}{d\eta}=105.21$ [Pa] with an error of only -1.77\%. This underscores the purpose of including this demonstration case -- even in cases with relatively large surface curvature like a cylinder, and cases with large pressure gradients such as in the strongly accelerating stagnation flow, where the magnitude of the streamwise gradient terms are on the order of or even larger than the viscous stress term, the relationship of Eq. \ref{eq:wall-shear-stress-balance-terms-annotated} still holds and meaningful shear stress estimates may be found.

\subsection{Mach 4.9, Smooth, Parameterized Curved-Wall, Turbulent Flow}
The final set of test cases comes from the NASA Langley Research Center Turbulence Modeling Resource \cite{NASA-LARC-Turb}. Specifically the DNS: High-Speed Turbulent Boundary Layers over Parameterized Curved Walls from Nicholson et al.~\cite{nicholson24}.~\footnote{Additional data to compute streamwise gradients was shared via private communication with Dhiman Roy and Lian Duan.} First, the ZPG (flat-plate portion of the surface) is analyzed; the forward-facing wall (FFW) and backward-facing wall (BFW) cases from \cite{nicholson24} are contained in the subsequent two subsections. For both curved wall cases, a portion of the respective surfaces have a favorable pressure gradient (FPG) and adverse pressure gradient (APG) portion associated with the convexity or concavity of the surface; for this demonstration we focus our attention to the region near the inflection point. For all, the $\alpha=1.0$ parameter is selected because the flow remains attached with this wall geometry.

\subsubsection{ZPG Flat Plate}
\begin{figure}[h!]
\centering
\captionsetup{justification=centering}
\begin{subfigure}{0.5\textwidth}
    \centering
    \captionsetup{justification=centering}
    \includegraphics[width=\textwidth]{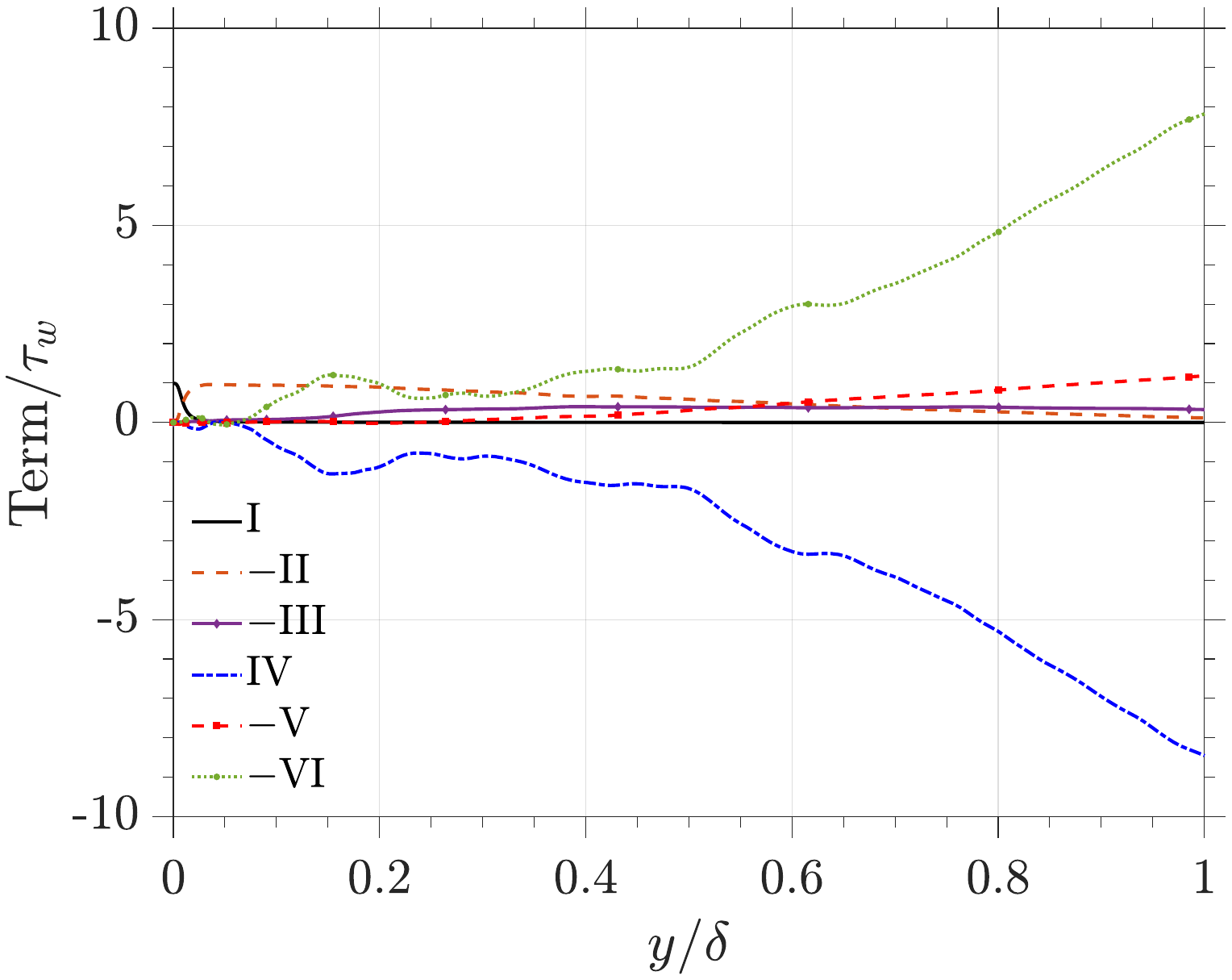}
    \caption{Contribution of stress term}
    \label{fig:M4p9-ZPG-bb-terms}
\end{subfigure}%
~
\begin{subfigure}{0.5\textwidth}
    \centering
    \captionsetup{justification=centering}
    \includegraphics[width=\textwidth]{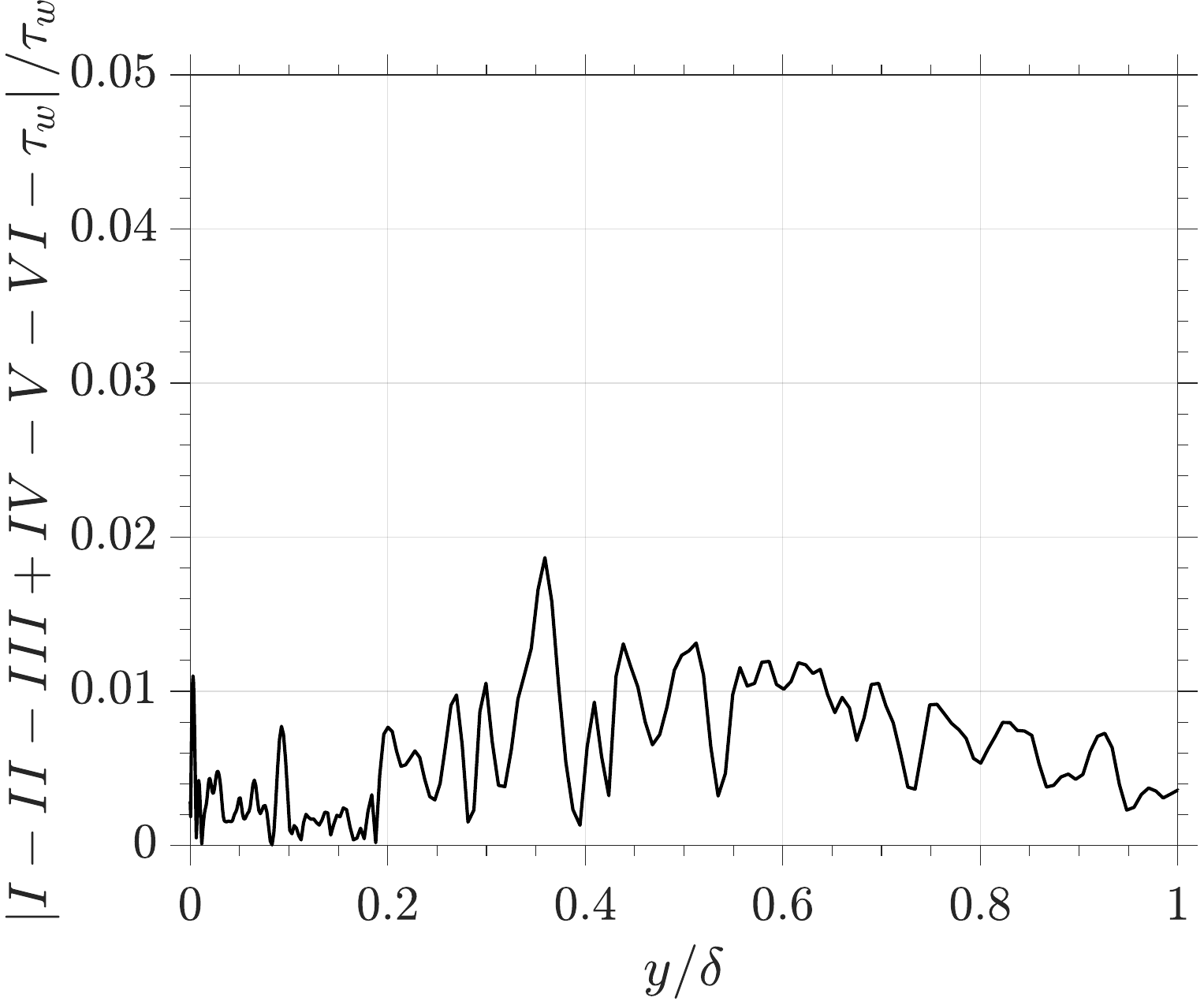}
    \caption{Error}
    \label{fig:M4p9-ZPG-bb-err}
\end{subfigure}
\caption{Mach 4.9, ZPG, smooth-wall, turbulent flat plate: contribution of each stress term and error compared to wall shear stress computed from velocity gradient at the wall as a function of wall-normal distance.}
\label{fig:M4p9-ZPG-terms}
\end{figure}

The first of the three parameterized curved wall demonstration cases is the flat plate portion of the wall just before the forward-facing slope. For clarity, in the original paper and NASA Langley resource, this location is labeled \textit{U2 forward-facing wall}. Figure \ref{fig:M4p9-ZPG-terms} shows the contribution of the stress terms and the local error at each wall-normal position when comparing the shear stress estimate from the sum of the terms, to the shear stress computed from the velocity gradient at the wall. Only the stress balance plots and the associated error are shown here; simulation details, flow profiles, and additional analysis can be found in the original work by Nicholson et al. \cite{nicholson24}. Figure \ref{fig:M4p9-ZPG-bb-terms} is similar to that from the Mach 2.5, ZPG, Laminar Flat Plate (section \ref{sec:Mach 2.5, ZPG, Laminar Flat Plate}) and the Mach 2.9, ZPG, Rough-Wall Turbulent Flat Plate (section \ref{sec:Mach 2.9, ZPG, Rough-Wall, Turbulent Flat Plate}). The streamwise gradient terms are similar in magnitude to the viscous and turbulent Reynolds stresses; therefore the contributions from each term are equally visible. For this case, the shear stress from the velocity gradient at the wall is 76.40 [Pa], and from the average of the momentum balance equation across the boundary layer is 76.56 [Pa] (average error of 0.21\%). Finally, from Fig. \ref{fig:M4p9-ZPG-bb-err}, the local error never exceeds 2\%. This demonstration case exemplifies most of the previous flat-plate examples and highlights all six non-negligible terms in Eq. \ref{eq:wall-shear-stress-balance-terms-annotated}.

\subsubsection{Forward-Facing Wall}
\begin{figure}[h!]
\centering
\captionsetup{justification=centering}
\begin{subfigure}{0.5\textwidth}
    \centering
    \captionsetup{justification=centering}
    \includegraphics[width=\textwidth]{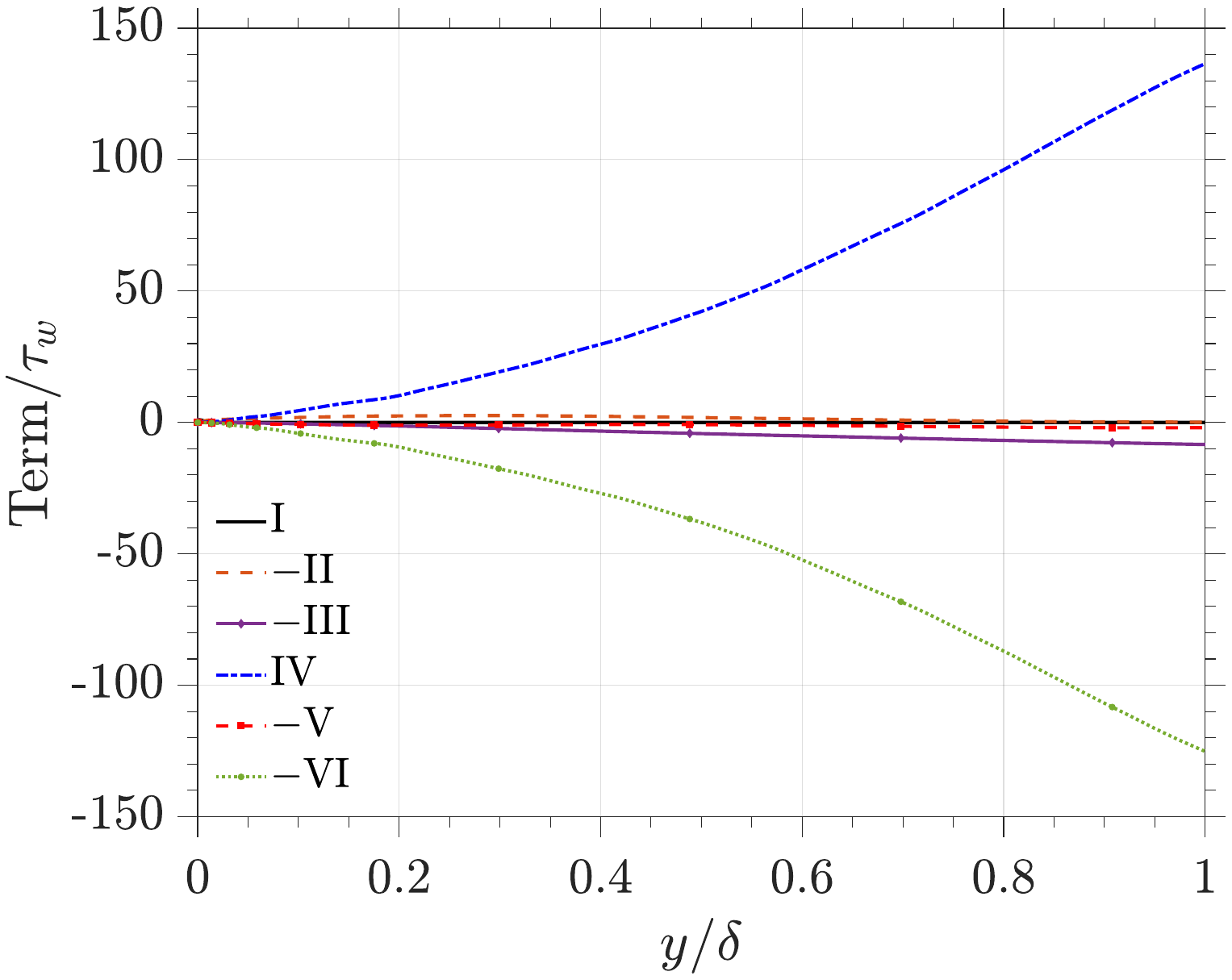}
    \caption{Contribution of stress term}
    \label{fig:M4p9-FFW-bb-terms}
\end{subfigure}%
~
\begin{subfigure}{0.5\textwidth}
    \centering
    \captionsetup{justification=centering}
    \includegraphics[width=\textwidth]{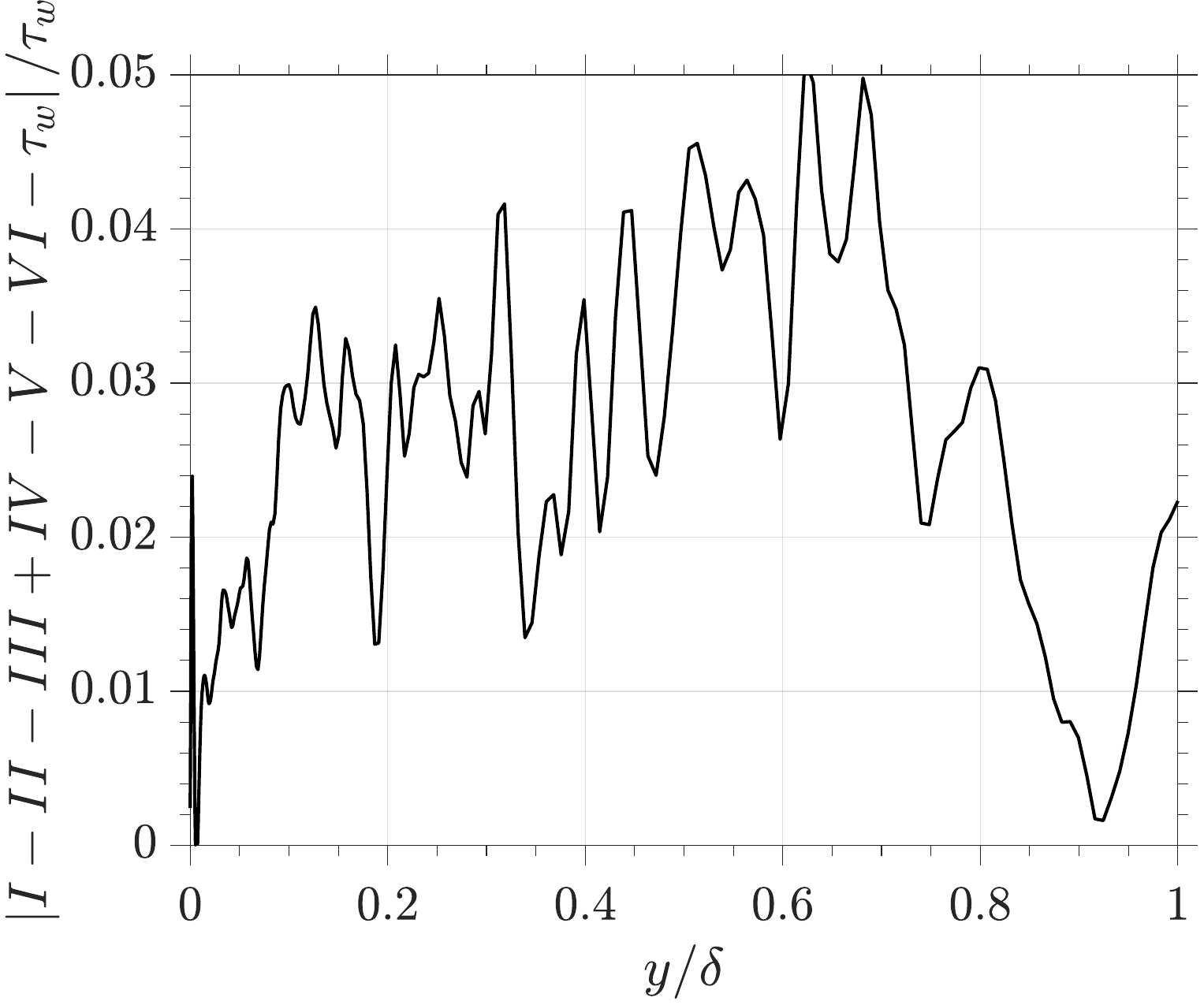}
    \caption{Error}
    \label{fig:M4p9-FFW-bb-err}
\end{subfigure}
\caption{Mach 4.9, smooth-wall, turbulent FFW: contribution of each stress term and error compared to wall shear stress computed from velocity gradient at the wall as a function of wall-normal distance.}
\label{fig:M4p9-FFW-terms}
\end{figure}

Continuing from the flat portion of the wall, the next station we consider is located in the middle of the forward-facing slope -- in the original work the location labeled \textit{L2 forward-facing wall}. It is again expected that all six non-negligible terms from Eq. \ref{eq:wall-shear-stress-balance-terms-annotated} will be active; however, the presence of the high surface curvature will make terms III, IV, V, and VI dominant. Due to the nature of the surface curvature, at the inflectional L2 location, the streamwise pressure gradient happens to be positive, but so does the streamwise velocity gradient; therefore this location was chosen in particular to emphasize the streamwise gradient contributions and their complex interactions. In particular, terms IV and VI become large immediately from the wall and by the edge of the boundary layer each have a magnitude over 100 times that of the actual wall shear stress. Given their outsized contribution (and subsequent cancellation of each other), accurately computing them is paramount. Figure \ref{fig:M4p9-dru_dx-FWD} shows the direct calculation of the streamwise mass flux gradient from a 4th order central difference in the streamwise direction, as compared to swapping them out with wall-normal gradients from the continuity equation, Eq. \ref{eq:continuity-gradient-swap}. Similar to previous discussion, eliminating the direct calculation of the streamwise gradient and replacing it with wall-normal gradients significantly improves the smoothness and avoids spurious oscillations that impact the shear stress estimate. To complete the discussion for this demonstration, the contribution of each stress term and the associated local error in the integral balance is shown in Fig. \ref{fig:M4p9-FFW-terms}. The wall shear stress from velocity gradient at wall at this location on the FFW is 116.10 [Pa], whereas the average wall shear stress from the momentum equation balance across the boundary layer is 113.67 [Pa]. Furthermore, the average percent error in the wall shear stress estimate is -2.09\%, and the local error remains below 5\% within the boundary layer.

\subsubsection{Backward Facing Wall}
\begin{figure}[h!]
\centering
\captionsetup{justification=centering}
\begin{subfigure}{0.5\textwidth}
    \centering
    \captionsetup{justification=centering}
    \includegraphics[width=\textwidth]{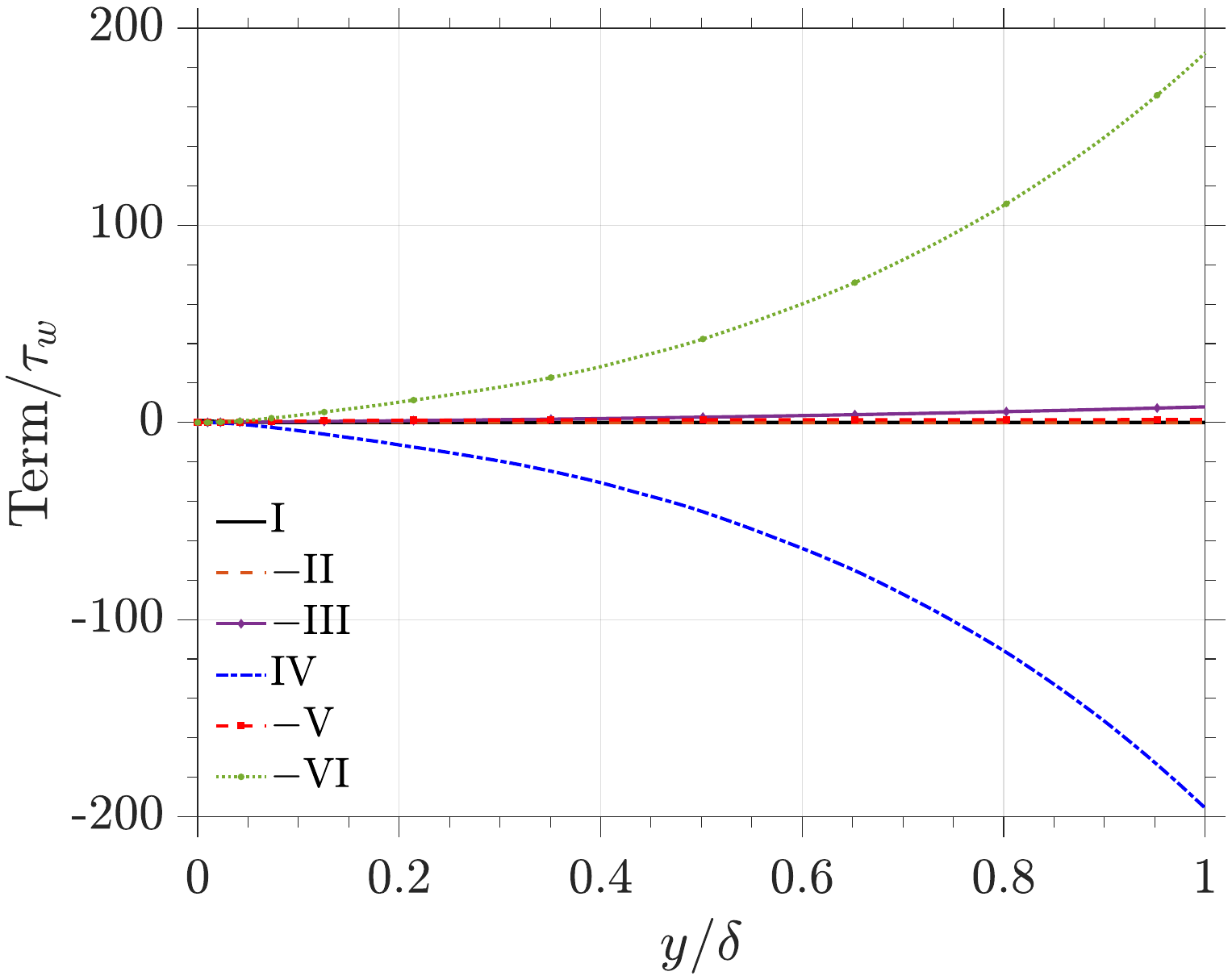}
    \caption{Contribution of stress term}
    \label{fig:M4p9-BFW-bb-terms}
\end{subfigure}%
~
\begin{subfigure}{0.5\textwidth}
    \centering
    \captionsetup{justification=centering}
    \includegraphics[width=\textwidth]{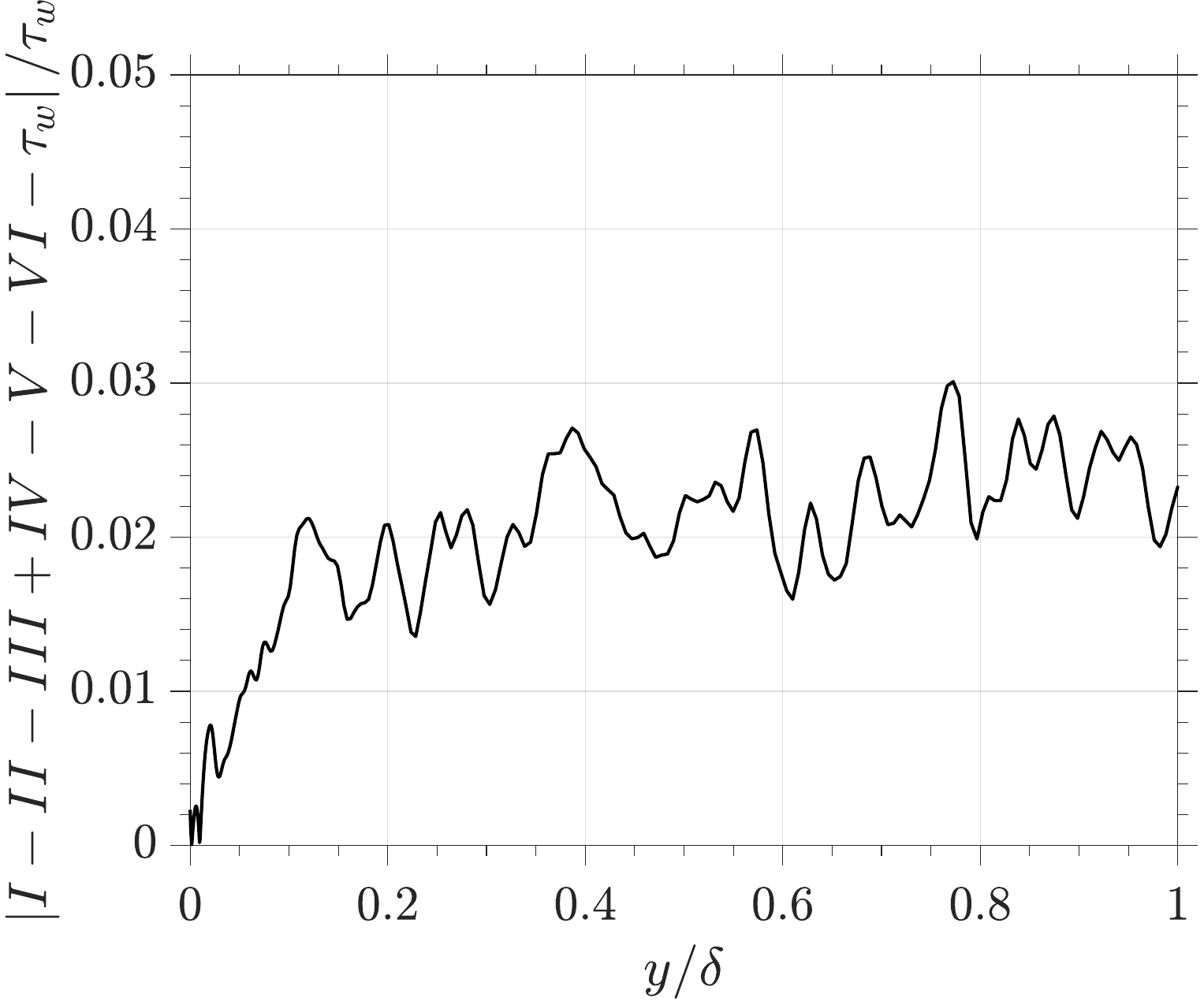}
    \caption{Error}
    \label{fig:M4p9-BFW-bb-err}
\end{subfigure}
\caption{Mach 4.9, smooth-wall, turbulent BFW: contribution of each stress term and error compared to wall shear stress computed from velocity gradient at the wall as a function of wall-normal distance.}
\label{fig:M4p9-BFW-terms}
\end{figure}

\begin{figure}[h!]
\centering
\captionsetup{justification=centering}
\begin{subfigure}{0.5\textwidth}
    \centering
    \captionsetup{justification=centering}
    \includegraphics[width=\textwidth]{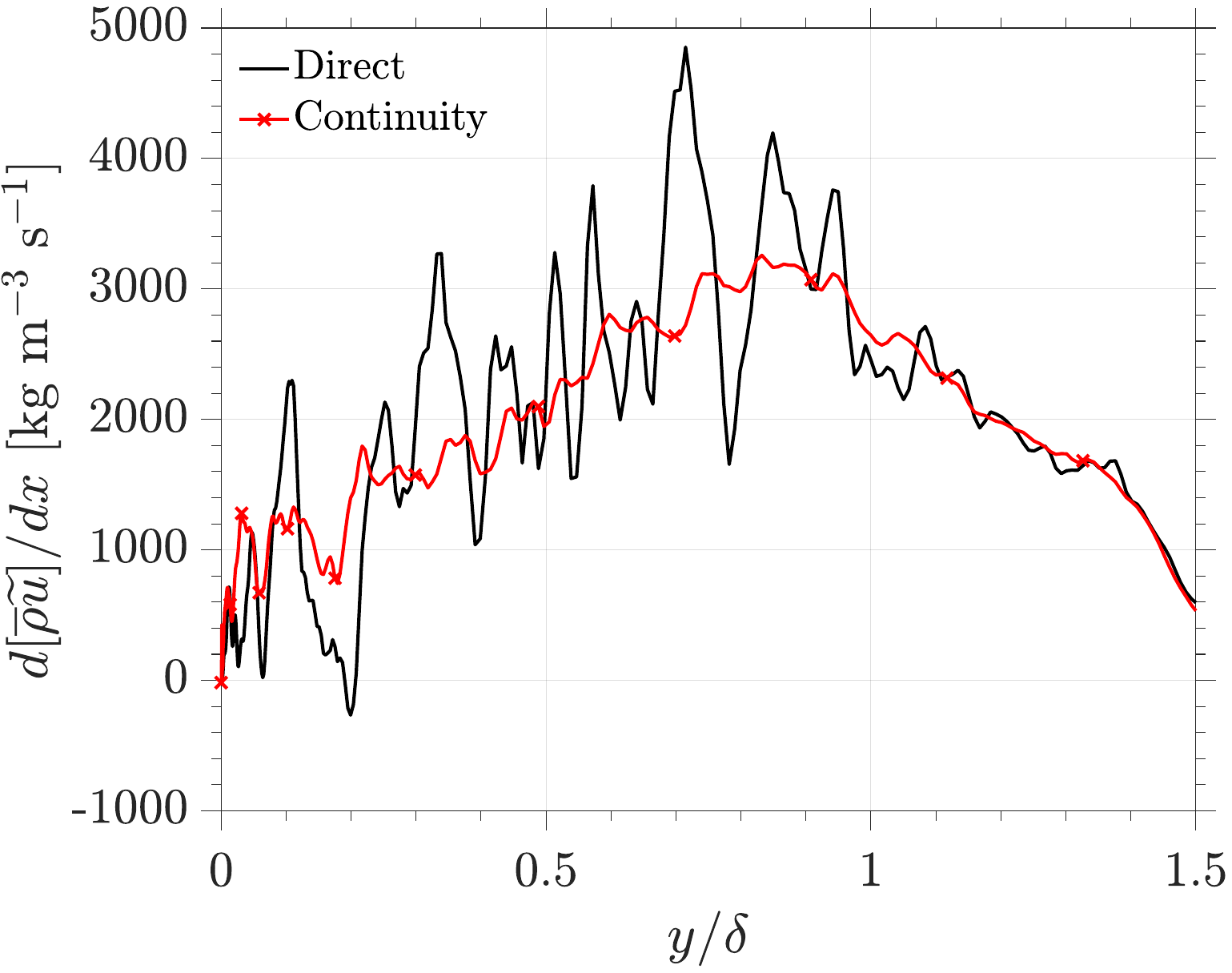}
    \caption{Forward-facing wall}
    \label{fig:M4p9-dru_dx-FWD}
\end{subfigure}%
~
\begin{subfigure}{0.5\textwidth}
    \centering
    \captionsetup{justification=centering}
    \includegraphics[width=\textwidth]{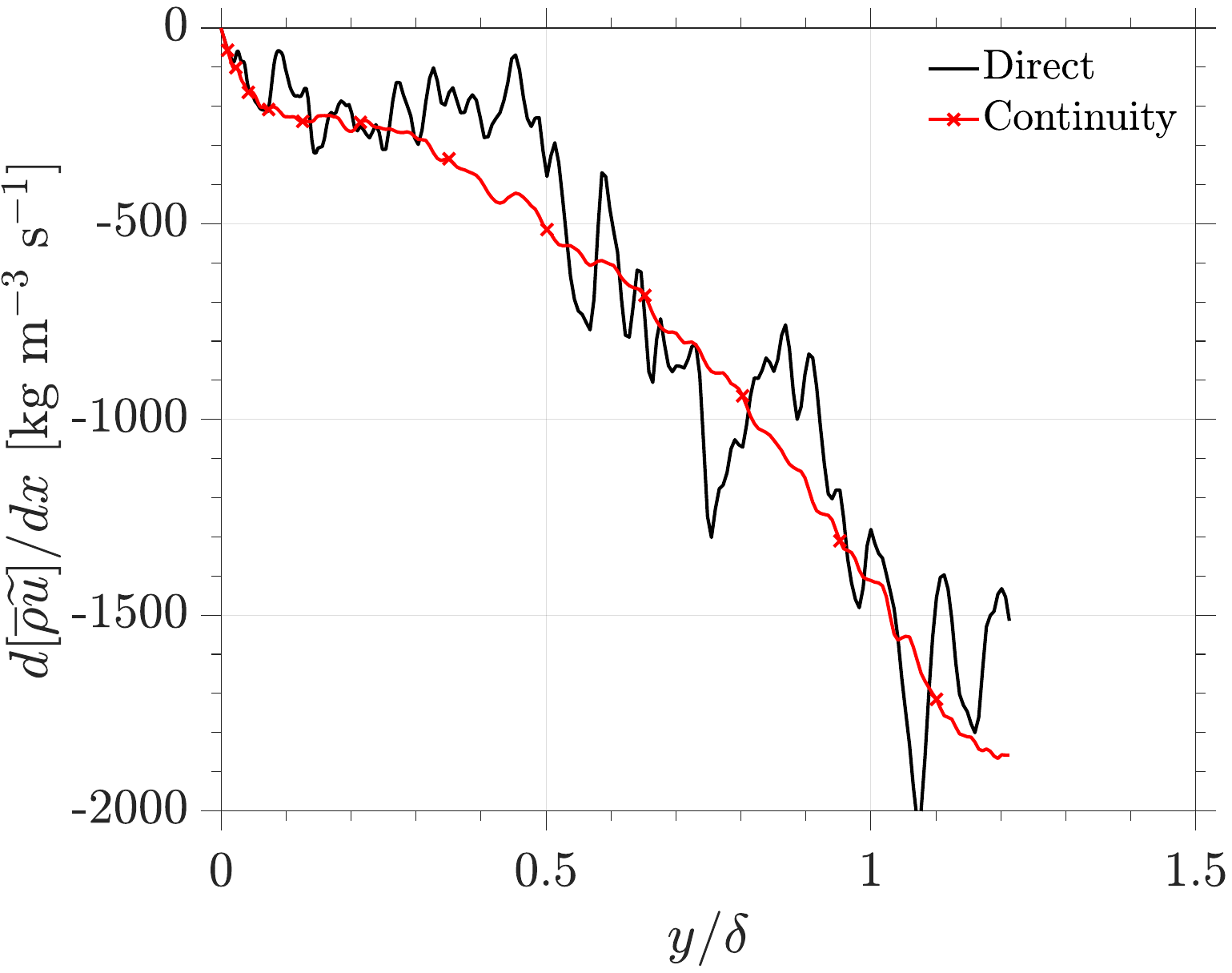}
    \caption{Backward-facing wall}
    \label{fig:M4p9-dru_dx-BWD}
\end{subfigure}
\caption{Mach 4.9, smooth-wall, turbulent, parameterized curved walls streamwise mass flux gradient comparison.}
\label{fig:M4p9-str-gradients}
\end{figure}

The last location analyzed is the middle of the backward facing slope -- in the original work the location labeled \textit{L2 backward-facing wall}.  Similar to the compression curve, all six non-negligible terms from Eq. \ref{eq:wall-shear-stress-balance-terms-annotated} are active; terms III, IV, V, and VI are dominant, and terms IV and VI become large immediately from the wall (albeit with opposite sign to the FFW). Figure \ref{fig:M4p9-dru_dx-BWD} shows the direct calculation of the streamwise mass flux gradient from a 4th order central difference in the streamwise direction, as compared to swapping them out with wall-normal gradients from the continuity equation, Eq. \ref{eq:continuity-gradient-swap}. Again, eliminating the direct calculation of the streamwise gradient and replacing it with wall-normal gradients significantly improves the smoothness and avoids spurious oscillations that impact the shear stress estimate.  Figure \ref{fig:M4p9-BFW-terms} shows the contribution of each stress term and the associated local error in the estimated wall shear stress as compared to that from the velocity gradient at the wall. The wall shear stress from the velocity gradient at the wall at this location on the BFW is 33.30 [Pa], whereas the average wall shear stress from the momentum equation balance across the boundary layer is 33.80 [Pa]. The average percent error in the wall shear stress estimate is 1.50\%, and the local error remains below 3\% within the boundary layer.

\subsection{Discussion}

A summary of the mean error in the shear stress estimate for each demonstration case is summarized in Table~\ref{tb:error_summary}. For all eight demonstration cases, the error between the stress balance estimate and a more traditional method is less than 5\%. This shows that the method is applicable regardless of whether the flow is turbulent, over a curved surface, over a rough surface, or has strong streamwise gradients. In practice, a useful metric to quantify the quality of the momentum integral shear stress estimate is to see the variation in the sum of the contributing terms across the boundary layer. A necessary but not sufficient condition for the estimate to reflect the true $\tau_w$ is that the stress balance produces a constant value throughout the boundary layer. If there are large oscillations or the sum of the terms increases or decreases significantly, then there are issues with the prediction.

The greatest challenge to an accurate estimate from this simple integral method are accurate streamwise gradients and subsequent integral terms III, IV, V, and VI. The gradient terms are sensitive to the quality of the available data and even with DNS data are subject to errors depending on the computational mesh resolution and averaging procedure. Moreover, if near-wall data is missing, the selection of the lower bound for the integrals in Eq. \ref{eq:wall-shear-stress-balance-terms-annotated} may introduce large error.  Nevertheless, the nature of the integral method permits a rough estimate of the shear stress by approximating the near-wall profiles and then integrating as usual once data becomes available. Eq. \ref{eq:low-up-integral} decomposes the integral terms into near-wall (low) and outer (up) integrals, where the first available data is at $y=y_C$: 

\begin{equation}
    \int_0^{y_P}  f(y) dy = \int_0^{y_C} f_{low}(y) dy + \int_{y_c}^{y_P} f_{up}(y) dy
    \label{eq:low-up-integral}
\end{equation}

If the near-wall data is missing the lower integrand $f_{low}(y)$ may be approximated. To illustrate the practicality, three approximations were considered: a constant approximation where $f_{low}=0$ or $f_{low}=f(y_C)$; a linear approximation with $f_{low}(y)=C_0y$, $C_0=f(y_C)/y_C$; and a quadratic fit to the available data $f_{low}=C_1y+C_2y^2$. Both the linear and quadratic approximation imply zero value for the integrand at $y=0$. It is not guaranteed that increasing order of the polynomial will improve the shear stress estimate; rather, it depends on the nature of the underlying streamwise gradient $\pp{}{x}$ profile. Observing the gradient profiles from the eight demonstration cases, often the gradients $\pp{[\overline{\rho}\widetilde{u}]}{x}$ and $\pp{\widetilde{u}}{x}$ go to zero at $y/\delta=0$, but $\pp{\overline{p}}{x}$ generally remains constant across the boundary layer and does not go to zero at the wall unless it is a ZPG boundary layer. Therefore, we recommend holding the pressure gradient approximation constant at the value of the first available pressure gradient data rather than using a fit that forces the profile through zero, $f_{low}(y)=\left.\pp{\overline{p}}{x}\right|_{y=y_C}$ . For the channel flow case, this means that the estimate for all three integral approximation methods will be the same because the only integral is the pressure term, and they will all have the same constant value of pressure gradient. 

In addition to the full data summary ($y_C/\delta=0.0$ column), Table \ref{tb:error_summary} also shows columns $y_C/\delta=0.2$ and $y_C/\delta=0.4$ to explore the performance of the near-wall integral approximations when 20\% and 40\% of the boundary layer data is missing. When fitting the quadratic, data from $y_C/\delta$, $1.1(y_C/\delta)$, $1.5(y_C/\delta)$, and $2.0(y_C/\delta)$ were used. For most cases, a rough approximation for the shear stress can be obtained within 10\% error for either a constant, linear, or quadratic approximation. The cases where the near-wall approximations did not work well are the curved wall cases. This is because the pressure gradient term does not remain constant across the boundary layer in the wall-normal direction and so holding it zero, holding it constant, or even low-order fits do not work well. For these conditions, caution should be used when approximating the underlying streamwise gradients, and alternative methods to infer the shear stress from normalized profiles like the incompressible formulation from Volino and Schultz \cite{Volino18} would be favorable because the near-wall approximation only depends on the inner scaled velocity $U^+$ and can be predicted easily. Additionally, the fits are sensitive to oscillations in the underlying data -- depending on the data points chosen for the fit, oscillations skewed the fit. Smoothing of the data may be a potential solution. However, caution should be used when approximating the near-wall integrals as well as smoothing of the data, because both do not guarantee the underlying conservation of mass and momentum from which the integral definition is based upon. Nonetheless, there is still value in the present method because with an approximation of the near-wall integral, an estimate of the shear stress may still be obtained; whereas traditional methods that require surface data or the velocity profile in the log-layer will fail to even provide a rough approximation. Ultimately, the benefit of the present formulation is in its simplicity by direct inference of the shear stress without iteration or normalization, and only requires the calculation of six contributing terms from wall-normal profiles.

\begin{sidewaystable}[p]
\caption{Summary of the error in the wall shear stress estimate from the stress balance method compared with the true value for the demonstration cases considered. Errors 0\% -- 5\% green, 5\% -- 12\% blue, $>$12\% red. The performance of the integral stress balance when near-wall data is missing is also included. ``--'' Indicates an erroneous negative shear stress value.}\label{tb:error_summary}
\centering
\scriptsize
\begin{tabular}{
    l
    S[table-format=3.2] 
    S[table-format=3.0] c 
    *{3}{S[table-format=3.0] @{\hspace{12pt}} c} 
    *{3}{S[table-format=3.0] @{\hspace{12pt}} c} 
}
\toprule
\multicolumn{1}{c}{Case} &
\multicolumn{1}{c}{\makecell{Ref. $\tau_w$  [Pa]}} &
\multicolumn{2}{c}{\makecell{Stress Bal.  [Pa]\\($y_C/\delta=0.0$)}} &
\multicolumn{6}{c}{\makecell{Stress Bal.  [Pa]\\($y_C/\delta=0.2$)}} &
\multicolumn{6}{c}{\makecell{Stress Bal.  [Pa]\\($y_C/\delta=0.4$)}} \\
\cmidrule(lr){3-4} \cmidrule(lr){5-10} \cmidrule(lr){11-16}
& &  & & \multicolumn{2}{c}{Const.} & \multicolumn{2}{c}{Lin.} & \multicolumn{2}{c}{Quad.}
& \multicolumn{2}{c}{Const.} & \multicolumn{2}{c}{Lin.} & \multicolumn{2}{c}{Quad.} \\
\cmidrule(lr){5-6} \cmidrule(lr){7-8} \cmidrule(lr){9-10}
\cmidrule(lr){11-12} \cmidrule(lr){13-14} \cmidrule(lr){15-16}
&  & {Val.} & {\% Err.} &
\multicolumn{1}{c}{Val.} & \multicolumn{1}{c}{\% Err.} &
\multicolumn{1}{c}{Val.} & \multicolumn{1}{c}{\% Err.} & \multicolumn{1}{c}{Val.} & \multicolumn{1}{c}{\% Err.} &
\multicolumn{1}{c}{Val.} & \multicolumn{1}{c}{\% Err.} &
\multicolumn{1}{c}{Val.} & \multicolumn{1}{c}{\% Err.} & \multicolumn{1}{c}{Val.} & \multicolumn{1}{c}{\% Err.} \\
\midrule
\makecell[l]{Mach 2.5, ZPG, Laminar,\\ Flat Plate [Present Work]} &
77.96 & 78.02 & \tcg{0.08} &
82.41 & \tcb{5.71} & 79.93 & \tcg{2.53} & 77.73 & \tcg{-0.30} 
& 86.26 & \tcb{10.65} & 93.05 & \tcr{19.36} & 87.03 & \tcb{11.63} \\\\
\makecell[l]{Mach 2.5, Smooth-Wall,\\Turbulent Channel Flow \cite{Gerolymos23,Gerolymos24,Gerolymos_database24}} &
269.73 & 257.91 & \tcg{-4.38} &
261.89 & \tcg{-2.91} & 261.89 & \tcg{-2.91} & 261.89 & \tcg{-2.91} 
& 263.50 & \tcg{-2.31} & 263.50 & \tcg{-2.31} & 263.50 & \tcg{-2.31} \\\\
\makecell[l]{Mach 4.0, Rough-Wall,\\Turbulent Channel Flow \cite{Modesti22,Modesti22_database}} &
185.91 & 185.78 & \tcg{-0.07} &
185.78 & \tcg{-0.07} & 185.78 & \tcg{-0.07} & 185.78 & \tcg{-0.07} 
& 185.78 & \tcg{-0.07} & 185.78 & \tcg{-0.07} & 185.78 & \tcg{-0.07} \\\\
\makecell[l]{Mach 2.9, ZPG, Rough-Wall,\\Turbulent Flat Plate [Present Work]} &
437.35 & 437.46 & \tcg{0.03} &
418.39 & \tcg{-4.34} & 467.97 & \tcb{7.00} & 437.12 & \tcg{-0.05} 
& 474.56 & \tcb{8.51} & 461.08 & \tcb{5.43} & 576.43 & \tcr{31.80} \\\\
\makecell[l]{Mach 6.0, Laminar, Hypersonic,\\Blunt Body [Present Work]} &
105.21 & 103.34 & \tcg{-1.77} &
116.65 & \tcb{10.87} & 103.80 & \tcg{-1.34} & 105.19 & \tcg{-0.02} 
& 146.83 & \tcr{39.56} & 106.28 & \tcg{1.02} & 108.24 & \tcg{2.88} \\\\
\makecell[l]{Mach 4.9, Smooth, ZPG Flat\\Plate, Turbulent Flow \cite{NASA-LARC-Turb,nicholson24}} &
76.40 & 76.56 & \tcg{0.21} &
109.47 & \tcr{43.29} & 120.74 & \tcr{58.04} & 126.62 & \tcr{65.73} 
& 62.27 & \tcr{-18.49} & 63.72 & \tcr{-16.60} & 78.15 & \tcg{2.29} \\\\
\makecell[l]{Mach 4.9, Smooth, Forward-Facing\\Wall, Turbulent Flow \cite{NASA-LARC-Turb,nicholson24}} &
116.10 & 113.67 & \tcg{-2.09} & \text{--} & \text{--} & 22.59 & \tcr{-80.54} & 69.28 & \tcr{-40.33} 
& \text{--} & \text{--} & \text{--} & \text{--} & \text{--} & \text{--} \\\\
\makecell[l]{Mach 4.9, Smooth, Backward-Facing\\Wall, Turbulent Flow \cite{NASA-LARC-Turb,nicholson24}} &
33.30 & 33.80 & \tcg{1.50} &
75.13 & \tcr{125.62} & 61.46 & \tcr{84.56} & 57.78 & \tcr{73.51} 
& 138.52 & \tcr{315.98} & 105.92 & \tcr{218.08} & 89.89 & \tcr{169.94} \\
\bottomrule
\end{tabular}
\end{sidewaystable}

\section{Conclusions} \label{sec:Conclusions}
Inspired by existing momentum integral equations to infer the wall shear stress, this work formalizes a simple integral equation by once-integrating the Favre-averaged streamwise momentum equation in the wall-normal direction to arrive at a stress balance consisting of six contributing terms. Additionally, we present substitutions for the problematic streamwise gradient terms such that they depend only on wall-normal profiles or wall-normal gradients, obtained through manipulation of the differential form of the governing continuity and momentum equations. 

Applying the integral relation as a post-processing step for scale-resolving data, we show the contributions of the various terms of the integral equation, the associated error in the estimate, and outline practical considerations when estimating the wall shear stress for complex flow conditions. The adequacy of this technique was demonstrated through eight test cases. The conditions tested are high-speed, compressible flows, in particular complex flows involving pressure gradients, surface roughness, and surface curvature. In all cases, the average predicted shear stress estimate from the present integral relation compared to a more traditional approach was no more than 5\%, with many cases having much lower error. In addition, the present integral formulation does not have a significant dependence on the upper limit of integration; rather, the important bound of integration is the lower bound from which the integral relation is valid up to any wall-normal location. For most cases, approximating the near-wall data with constant, linear, or quadratic profiles produced acceptable estimates of shear stress even in the extreme case when 40\% of the boundary layer data was missing; although, the method struggles when the pressure gradient does not remain constant across the boundary layer or there are significant oscillations in the underlying data. Nevertheless, we found the method to be viable even for surfaces in the transitionally rough and full rough regimes by appropriate selection of a virtual origin. Ultimately, the method demonstrated its ability to represent a variety of flow conditions, and the approach was found to be general and applicable regardless of surface curvature, streamwise gradients, or surface roughness.

\begin{acknowledgments}
This work has been supported under a NASA Space Technology Research Institute Award (ACCESS, grant number 80NSSC21K1117). The authors would like to thank Graham V. Candler and the University of Minnesota for their computing resources. The authors would also like to thank Dhiman Roy and Lian Duan from The Ohio State University for sharing additional Mach 4.9, Smooth, Parameterized Curved-Wall, Turbulent Flow data.

\end{acknowledgments}


\bibliography{apssamp}

\end{document}